\documentclass[final]{siamart}

\usepackage[capitalize]{cleveref}

\crefformat{section}{#2\S#1#3}
\crefrangeformat{section}{#3\S\S#1#4--#5#2#6}
\crefmultiformat{section}{#2\S\S#1#3}{ and #2#1#3}{, #2#1#3}{, and #2#1#3}
\crefrangemultiformat{section}{#3\S\S#1#4--#5#2#6}%
{ and #3#1#4--#5#2#6}{, #3#1#4--#5#2#6}{, and #3#1#4--#5#2#6}

\Crefformat{section}{#2Section~#1#3}
\Crefrangeformat{section}{#3Sections~#1#4--#5#2#6}
\Crefmultiformat{section}{#2Sections~#1#3}{ and #2#1#3}{, #2#1#3}{, and #2#1#3}
\Crefrangemultiformat{section}{#3Sections~#1#4--#5#2#6}%
{ and #3#1#4--#5#2#6}{, #3#1#4--#5#2#6}{, and #3#1#4--#5#2#6}

\crefformat{equation}{#2(#1)#3}
\crefrangeformat{equation}{(#3#1#4--#5#2#6)}
\crefmultiformat{equation}{#2(#1)#3}{ and #2(#1)#3}
{, #2(#1)#3}{, and #2(#1)#3}
\crefrangemultiformat{equation}{(#3#1#4--#5#2#6)}%
{ and (#3#1#4--#5#2#6)}{, (#3#1#4--#5#2#6)}{, and (#3#1#4--#5#2#6)}

\Crefformat{equation}{#2Equation~(#1)#3}
\Crefrangeformat{equation}{Equations~(#3#1#4--#5#2#6)}
\Crefmultiformat{equation}{Equations~(#2#1#3)}{ and (#2#1#3)}
{, (#2#1#3)}{, and (#2#1#3)}
\Crefrangemultiformat{equation}{Equations~(#3#1#4--#5#2#6)}%
{ and (#3#1#4--#5#2#6)}{, (#3#1#4--#5#2#6)}{, and (#3#1#4--#5#2#6)}

\usepackage{color}
\definecolor{red}{rgb}{1.0,0.,0.}
\definecolor{blue}{rgb}{0.,0.,1.0}
\definecolor{bluegreen}{rgb}{0.0,0.5,0.5}
\definecolor{brown}{rgb}{1.0,0.5,0.25}

\usepackage{amsmath}
\usepackage{amssymb}
\usepackage{multicol}

\usepackage{verbatim}
\usepackage{accents}
\usepackage{graphicx}
\usepackage{subfig}
\usepackage[section]{placeins}
\usepackage{multirow}

\newcommand{\ignore}[1]{}

\newenvironment{mat}{\left[ \begin{array}{ccccccccccccc}}{\end{array}\right]}
\newenvironment{rmat}{\left[
\begin{array}{rrrrrrrrrrrrr}}{\end{array}\right]}
\newenvironment{lmat}{\left[
\begin{array}{lllllllllllll}}{\end{array}\right]}
\newcommand\bcm{\begin{mat}}
\newcommand\ecm{\end{mat}}
\newcommand\brm{\begin{rmat}}
\newcommand\erm{\end{rmat}}
\newcommand\blm{\begin{lmat}}
\newcommand\elm{\end{lmat}}

\newcommand{\lame}{\lambda}
\newcommand{\sige}[2]{\sigma^{#1 #2}}
\newcommand{\vel}[1]{u^#1}

\graphicspath{.}

\begin{document}

\title{A high-resolution Finite Volume Seismic Model to Generate
Seafloor Deformation  for Tsunami Modeling}

\author{Christopher J. Vogl%
  \thanks{Department of Applied Mathematics, University of Washington,
Seattle, WA \email{chris.j.vogl@gmail.com}.}%
  \and
  Randall J. LeVeque%
  \thanks{Department of Applied Mathematics, University of Washington,
Seattle, WA \email{rjl@uw.edu}.}%
}

\maketitle


\pagestyle{myheadings}
\thispagestyle{plain}
\markboth{C. J. Vogl and R. J. LeVeque}%
{Finite Volume Seismic Model}

\begin{abstract}
A high-resolution finite volume method approach to incorporating time-dependent slip across rectangular subfaults when modeling general fault geometry is presented.  The fault slip is induced by a modification of the Riemann problem to the linear elasticity equations across cell interfaces aligned with the subfaults.  This is illustrated in the context of the high-resolution wave-propagation algorithms that are implemented in the open source Clawpack software ({\tt www.clawpack.org}), but this approach could be easily incorporated into other Riemann solver based numerical methods.  Surface deformation results are obtained in both two and three dimensions and compared to those given by the steady-state, homogeneous half-space Okada solution.
\end{abstract}

\begin{keywords}
Clawpack, GeoClaw, fault slip, Riemann problem, Okada, tsunami source, subduction zone, embedded interface
\end{keywords}


\section{Introduction}
\label{sec:intro}

Tsunamigenic megathrust earthquakes generally occur in subduction zones, where an oceanic plate is subducting beneath a continental plate.  Stress can build up in locked regions and be violently released.  The resulting seafloor motion causes a disturbance of the sea surface that gives rise to gravity waves in the fluid layer.  For recent events, it is often the slip on the fault surface that is estimated via inversion techniques from seismic or other observations \cite{hartzell_1994_1996,ji_source_2002,melgar_kinematic_2015}.  In order to perform tsunami simulation, it is necessary to transform this fault slip into seafloor deformation.  To do this, the earth is typically modeled as a homogeneous, isotropic half-space, where a Green's function solution exists for steady-state linear elasticity with a delta function displacement at a point in the interior.  Often the fault surface is approximated by a collection of planar subfaults, e.g. rectangular patches on which the slip is assumed to be constant.  Integrating the Green's function over a rectangular subfault gives an explicit expression for the surface deformation as a function of the parameters defining the subfault geometry and the slip.  By linearity, these can be summed over the subfaults in order to approximate the deformation from a more complex source \cite{titov_real-time_2005}.  In the tsunami modeling literature, this is often called the Okada solution, using the explicit formulas derived by Okada \cite{Okada1985} for rectangular patches.
\par
This solution, however, only approximates the final static deformation of the surface and does not approximate the transient motion that occurs as seismic waves propagate during the earthquake.  It also assumes the surface is flat, while in practice the seafloor is not flat.  In particular, the region of maximum surface displacement for subduction zone earthquakes is typically near the relatively steep continental slope at the edge of the shelf, which often terminates in a trench that is deeper than the ocean farther offshore.  In tsunami modeling, the Okada solution on the flat surface is typically transferred directly to the actual topography.  It is also often assumed that the water column above each point on the seafloor is instantaneously lifted by the sea floor deformation \cite{nosov_tsunami_2014}, causing a displacement of the initially flat sea surface that exactly matches the Okada solution.  In this case, one can think of the Okada model as incorporating the ocean into the elastic half space, ignoring the jump in material properties that occurs at the seafloor/water interface.
\par
Moreover, it is often assumed that the steady state static deformation occurs instantaneously during the earthquake, when in reality the rupture may grow and propagate over the course of several minutes.  This can be modeled to some extent using the Okada model by allowing subfaults to rupture at different times and by assuming the resulting static deformation grows over some ``rise time''
associated with the subfault. This quasi-static approach is often used in tsunami modeling,
particularly for events such as the 2004 Sumatra-Andaman earthquake that gave rise to the devastating Indian Ocean tsunami.  This 1200km long fault ruptured over the course of about 10 minutes.  The quasi-static Okada approach is
considered generally adequate for many modeling problems, taking into account the
uncertainty in the input variables (slip, rupture time, rise time) that are often poorly
constrained from observations and inversions, even for recent event.  This approach is
implemented in the GeoClaw software that is widely used for tsunami modeling \cite{BergerGeorgeLeVequeMandli:awr11,LeVequeGeorgeBerger:an11}, as well as by
Dutykh et al. \cite{dutykh_use_2013}.
\par
For some problems, however, more accurate estimates of sea surface
displacement may be required, and in some cases even the transient seismic
waves in the earth and associated acoustic waves in the ocean may be
important to model. For example, Dutykh and Dias \cite{dutykh_tsunami_2009}
noted that transient effects can cause a leading depression wave where an
elevation wave is expected.  An important potential application
is to the study of early warning systems
that might be used to provide enhanced warnings based on observations
obtained in the source region during and immediately following a major
earthquake \cite{nosov_tsunami_2014}.
Recent work by Kozdon and Dunham \cite{KozdonDunham2014Hydro},
by Maeda, Furumuro, and collaborators
\cite{MaedaFurumura2011,MaedaFurumuraEtAl2013}, and by
Saito and Tsushima \cite{SaitoTsushima2016}, for example,
 has shown that the study of coupled seismic and acoustic waves could be
useful in rapid inversion to distinguish between different rupture models.
Our work is a part of an ongoing study of the feasibility and potential uses of a cabled network of sea floor sensors to monitor the Cascadia Subduction Zone \cite{eew-whitepaper}.  Thus, the goal herein is to develop a high-resolution finite volume method for solving the seismic wave equations (eventually to be coupled with the ocean) that allows the specification of arbitrary slip on a fault surface and works well in the context of the open source Clawpack software \cite{clawpack,mandli2016clawpack}, which includes adaptive mesh refinement (AMR) algorithms in both two and three space dimensions. In addition to providing an efficient and accurate solver for these equations, this will facilitate coupling with the GeoClaw software that is also built on the Clawpack AMR framework and distributed as part of that software.
\par
This paper focuses on the homogeneous half-space problem in order to verify that the method and implementation presented here can reproduce the exact Okada solution.  After deriving the modified Riemann solutions in Sect.~\ref{sec:rp}, computed surface deformations are obtained for a two-dimensional vertical cross section and for a full three-dimensional problem in Sect.~\ref{sec:comparison}.  Concluding remarks and discussion of ongoing extensions of this work, including the addition of an ocean layer and seafloor topography, are found in Sect.~\ref{sec:conclusion}.  The code producing the simulations shown in this paper is archived at \cite{seismic-zenodo} and active development work can be followed in the  Github repository at {\tt github.com/clawpack/seismic}.

\section{The elasticity Riemann solver with fault slip}
\label{sec:rp}

The equations of isotropic linear elasticity can be written as
\begin{equation}\label{eq:3delasticity}
	\begin{split}
		&\sigma_t - \lambda (\nabla \cdot u)I - \mu (\nabla u + \nabla^T u) = 0,\\
		&\rho u_t - \nabla \cdot \sigma = 0,\\
	\end{split}
\end{equation}
where $\sigma$ is the symmetric stress tensor, $u$ is the velocity, $I$ is the identity tensor, $\rho$ is the the density, and $\lambda$, $\mu$ are the Lam\'e parameters.  The wave-propagation algorithm discussed in \cite{fvmhp} uses the solutions to the Riemann problems at grid cell interfaces to update cell quantities.  This approach extends to other materials, such as orthotropic and poroelastic materials \cite{Lemoine:3d,LemoineOu2014,LemoineOuLeVeque2013}.  Here, the fault slip is introduced by modifying the Riemann problems, and corresponding solutions, at cell interfaces that line up with the fault.  Given that two-dimensional simulations can be performed in much less time than in three dimensions, the plane-strain case is used first for comparing against the Okada solution for various faults of infinite length in the strike direction.  The slip model is also extended to three dimensions and compared to the corresponding Okada solution for a fault of finite length.

\subsection{Two dimensions (plane-strain)}\label{sec:2d}

In the two-dimensional, plane-strain case, the equations take the form
\begin{equation}\label{eq:2delasticity}
	\begin{split}
		&\sige11_t - (\lame+2\mu) \vel1_x - \lame \vel2_y = 0,\\
		&\sige22_t - \lame \vel1_x  - (\lame+2\mu) \vel2_y= 0,\\
		&\sige12_t - \mu(\vel2_x+\vel1_y) =0,\\
		&\rho \vel1_t - \sige11_x -\sige12_y =0,\\
		&\rho \vel2_t - \sige12_x -\sige22_y =0,
	\end{split}
\end{equation}
where superscripts denote components in the $x$ (horizontal) and $y$ (vertical) directions, respectively.  This models a vertical slice of the earth ($y \le 0$) with an infinitely long fault in the orthogonal direction.  In heterogeneous media, the density $\rho$ and the Lam\'e parameters $\lame$, $\mu$ can be spatially varying, which results in the following linear hyperbolic system of equations in non-conservative form:
\begin{equation*}
	q_t + A(x,y)q_x + B(x,y)q_y = 0,
\end{equation*}
where $q = [\sige11,\sige22,\sige12, \vel1,\vel2]^T$ and $A$ and $B$ are the $5 \times 5$ coefficient matrices.  As shown in \cite{fvmhp}, the general Riemann solution involves the eigenvectors of $n_x A + n_y B$, where $n = [n_x,n_y]^T$ is the normal vector to the cell-edge.  These vectors are
\begin{equation}\label{eq:waves}
	r^p_\pm = \bcm \lambda + 2 \mu n_x^2 \\ \lambda + 2\mu n_y^2 \\ 2\mu n_x n_y \\ \mp c_p n_x \\ \mp c_p n_y \ecm \quad \text{and} \quad
	r^s_\pm = \bcm - 2\mu n_x n_y \\ 2 \mu n_x n_y \\ \mu (n_x^2 - n_y^2) \\ \pm c_s n_y \\ \mp c_s n_x \ecm,
\end{equation}
where $r^p_\pm$ and $r^s_\pm$ correspond to the P- and S- waves traveling either in the direction of the normal ($+$) or opposite ($-$).
\par
Denote $q^*_\pm$ and $q_\pm$ as the initial and resulting states, respectively, on corresponding sides of the cell-edge.  These are related via
\begin{equation*}
	\begin{split}
		q_+ = q^*_+ - \alpha^p_+r^p_+ - \alpha^s_+ r^s_+, \\
		q_- = q^*_+ + \alpha^p_- r^p_- + \alpha^s_- r^s_-.
	\end{split}
\end{equation*}
where  the $\alpha$'s are amplitudes of the P-waves and S-waves.  To couple $q_+$ with $q_-$, continuity of normal traction ($\sigma_n := \sigma.n\cdot n$) and normal velocity ($u_n := u \cdot n$) are enforced across the cell-edge.  The fault slip is now implemented by enforcing a slip rate $s$ evenly to the two tangential velocities ($u_\tau$) on either side of the cell-edge.  These conditions are summarized in matrix form:
\begin{equation*}
	\begin{split}
		&P_n q_- = P_n  q_+, \quad P_\tau q_- = s/2, \text{ and }P_\tau q_+ = -s/2,\\
		&\text{where } P_n q :=  \bcm n_x^2 & n_y^2 & 2n_x n_y & 0 & 0 \\ 0 & 0 & 0 & n_x & n_y \ecm q  = \bcm \sigma _n \\ u_n \ecm \\
		&\text{ and } P_\tau q := \bcm 0 & 0 & 0 & n_y & -n_x \ecm q = \bcm u_\tau \ecm.
	\end{split}
\end{equation*}
The resulting states $q_\pm$ are exchanged for the initial states and eigenvectors to obtain equations for $\alpha^p_\pm$ and $\alpha^s_\pm$.  Noting that $P_n r^s_\pm = 0$ and $P_\tau r^p_\pm = 0$ helps to greatly simplify the system:
\begin{equation*}
	\begin{split}
		&\bcm P_n r^p_+ & P_n r^p_- & 0 & 0 \ecm \bcm \alpha^p_+ \\ \alpha^p_-\ecm = P_n  (q^*_+ - q^*_-),\\
		&\bcm P_\tau r^s_+ & 0 \\ 0 &  P_\tau r^s_- \ecm \bcm \alpha^s_+ \\ \alpha^s_- \ecm = \bcm P_\tau q^*_+ + s/2 \\ -P_\tau q^*_- + s/2 \ecm.
	\end{split}
\end{equation*}
The solution is
\begin{equation}\label{eq:rpsolution}
	\alpha^p_\pm = \frac{c_{p\mp}\Delta \sigma^*_n \mp B_\mp \Delta u^*_n}{c_{p+}B_- + c_{p-}B_+} \quad \text{and} \quad \alpha^s_\pm = \frac{u^*_{\tau\pm} \pm s/2}{c_{s\pm}},
\end{equation}
where $\Delta \sigma^*_n := \sigma^*_{n+} - \sigma^*_{n-}$ and $\Delta u^*_n := u^*_{n+} - u^*_{n-}$.
\par
At cell-edges that correspond to a subfault, the goal is to obtain a total displacement $S$ over a time period $\Delta t$, beginning at some rupture time $t_r$.  Thus, for time $t_r \leq t \leq t_r+ \Delta t$, the slip rate is imposed as $s = S/\Delta t$ (uniform slip in time).  The variable time-stepping algorithm in Clawpack is adjusted so that $\Delta t$ is divided into an integer number of time steps.  When the slip rate is zero, the standard Riemann solution is used where continuity of traction and velocity are enforced across the cell-edge.

\subsection{Additional numerical details}

\begin{figure}[t]
	\center
	\subfloat[$t = 0$s]{\includegraphics[width=0.49\textwidth]{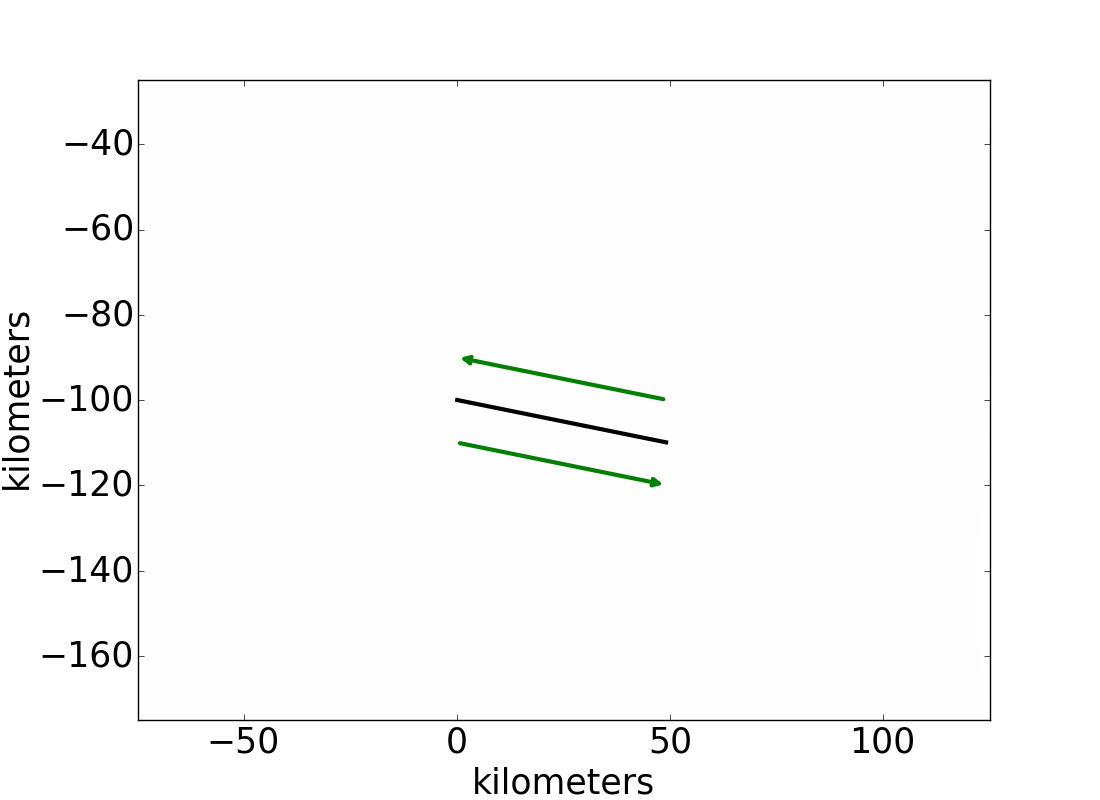}}
	\subfloat[$t = 1.5$s]{\includegraphics[width=0.49\textwidth]{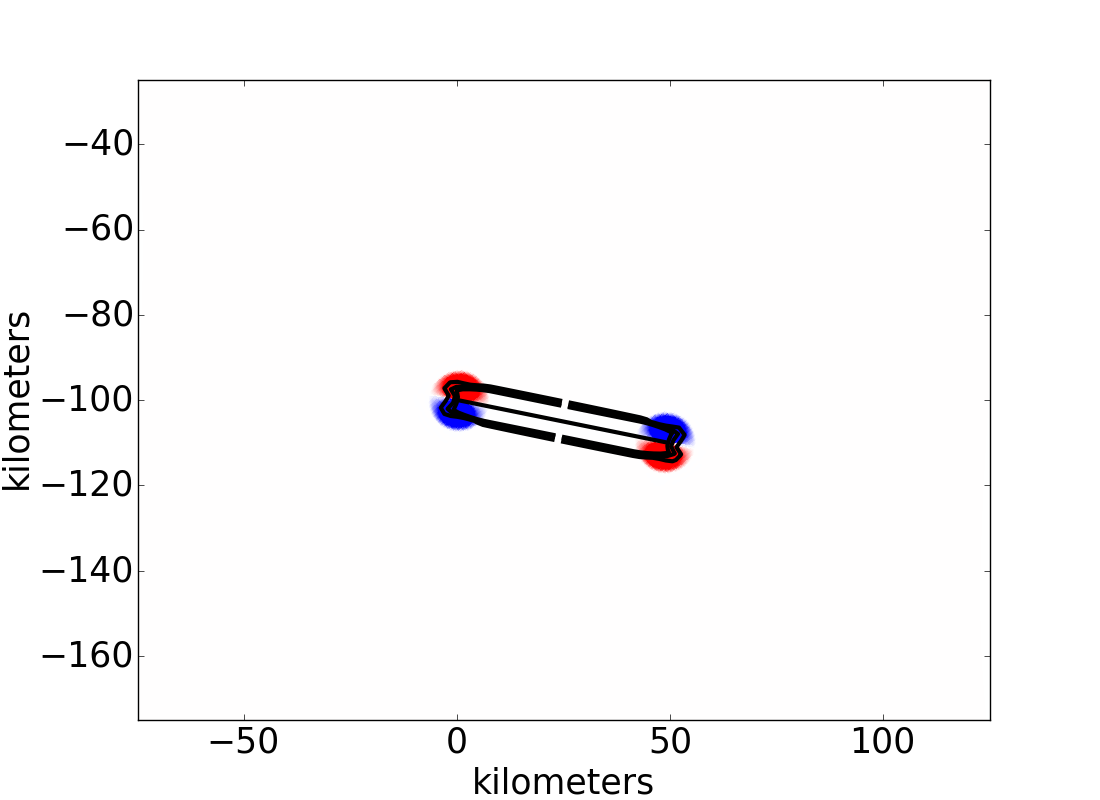}}\\
	\subfloat[$t = 5$s]{\includegraphics[width=0.49\textwidth]{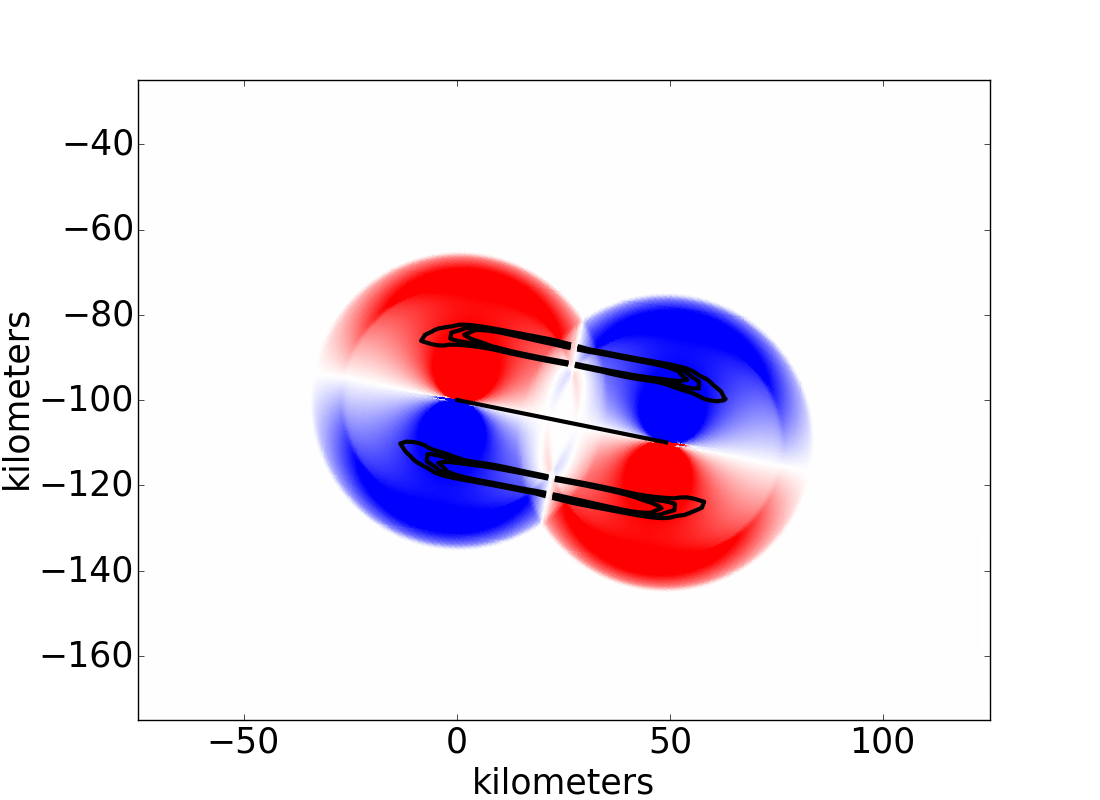}}
	\subfloat[$t = 10$s]{\includegraphics[width=0.49\textwidth]{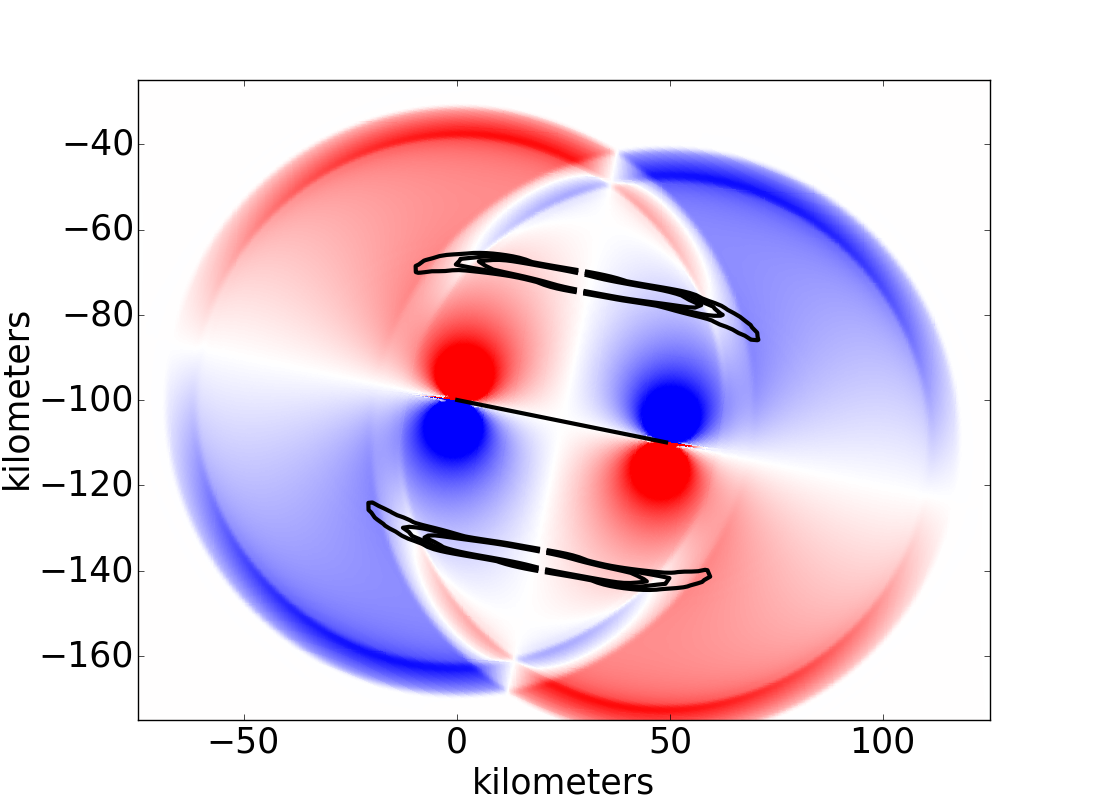}}
	\caption[Compression/tension and shear waves emanating from a dipping
fault]{Compression/tension and shear waves emanating from a $1$km, $1$s uniform slip across a fault with dip $0.2$rad, top-edge depth $100$km, and width $50$km, shown at four different times  (\textcolor{red}{red}: compression,
\textcolor{blue}{blue}: tension, black lines: contours of velocity in the direction tangent to the fault).}
	\label{fig:faultwaves}
\end{figure}

Fig.~\ref{fig:faultwaves} shows a few time frames of a simulation in which unit slip is imposed for $0 \le t \le 1$ across a fault with dip angle of $0.2$ radians ($\approx 11.5$ degrees), top-edge depth $100$km, and fault plane width $50$km.  The AMR capability of Clawpack is utilized with $8$ cells spanning the fault at the coarsest grid level and a Courant number set to $0.9$.  There are then 5 additional levels of grids, each twice as refined as the previous level in both space and time.  Note that as the seismic waves propagate out, a static stress field remains near the fault plane that indicates a permanent deformation of equal magnitude, but opposite sign, on either side of the fault.  For this figure, the fault is deep beneath the surface and so the waves continue propagating outward.  In reality, the depth of the fault may be quite shallow relative to its width, so the upward propagating waves would reflect off the surface ($y=0$) by the time shown in this figure.
\par
\begin{figure}[t]
	\center
	\includegraphics[width=0.40\textwidth]{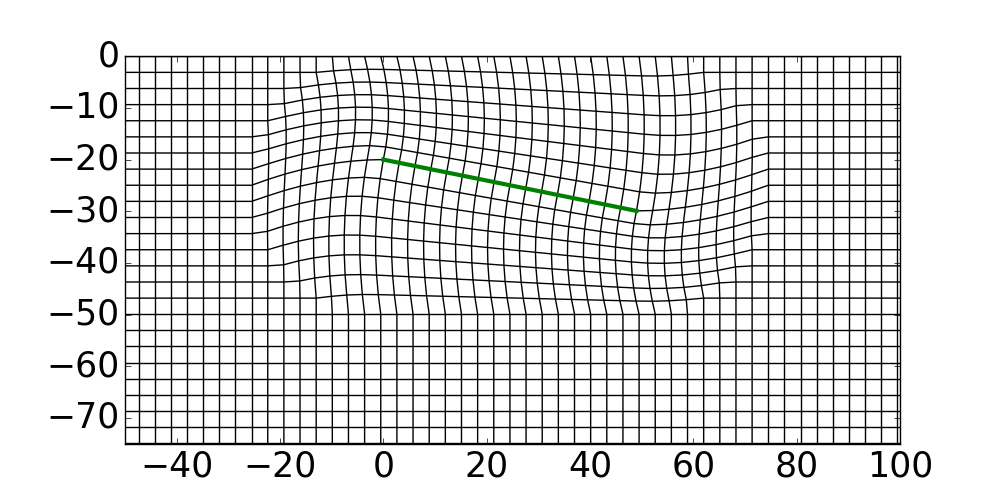}
	\caption{Mapped grid conforming both to fault and to surface.}
	\label{fig:mappedgrid}
\end{figure}

To address this, a free-surface boundary condition is imposed at the surface by setting ghost cell values in such a way that there is zero traction at the free surface.  More specifically, for $h>0$, the value of $\sige i2 (x,h)$ is set to $-\sige i2(x,-h)$, so that $\sige i2 (x,0) \approx  (1/2)[\sige i2 (x,h) + \sige i2 (x,-h)] = 0$ for $i=1,2$.  A mapped grid, $(X,Y) = \mathcal{M}(x,y)$, is used that lines up with both the fault and the free surface.  This is accomplished by interpolating between a normal Cartesian grid ($\mathcal{M}^s(x,y)=(x,y)$) and a grid that is rotated to line up with the fault: $\mathcal{M}^f (x,y) = (x_c + \cos(\theta)(x - x_c) + \sin(\theta)(y-y_c), y_c - \sin(\theta)(x - x_c) + \cos(\theta)(y-y_c))$, where $(x_c,y_c)$ is the centroid of the fault and $\theta$ the dip angle.  A distance from the fault is defined as
\begin{equation*}
	\phi(x,y) = \left \{ \begin{tabular}{cl}
		$\sqrt{(x - x_l)^2 + (y-y_c)^2}$ & for $x < x_l$ \\
		$|y-y_c|$ & for $x_l \le x \le x_r$ \\
		$\sqrt{(x - x_r)^2 + (y-y_c)^2}$ & for $x_r < x$ \end{tabular} \right .,
\end{equation*}
where $x_l = x_c - W/2$ and $x_r = x_c + W/2$ for fault width $W$.  The grid mapping is now complete as $\mathcal{M}(x,y) = \mathcal{M}^s(x,y)\phi(x,y)/y_c + \mathcal{M}^f(x,y)(1 - \phi(x,y)/y_c)$, for $\phi(x,y) \le y_c$, and $\mathcal{M}(x,y) = \mathcal{M}^s(x,y)$ otherwise.  A visualization of this mapped grid is found in Fig.~\ref{fig:mappedgrid}.

\subsection{Extension to three dimensions}\label{sec:3d}

Consider the three-dimensional elasticity equation (\ref{eq:3delasticity}) written in non-conservative form:
\begin{gather*}
	q_t + A(x,y,z)q_x + B(x,y,z)q_y + C(x,y,z)q_z = 0, \\
	q = [\sige11, \sige22, \sige33, \sige12, \sige23, \sige13, \vel1, \vel2, \vel3]^T,
\end{gather*}
where positive $x$ is east, positive $y$ is north, and positive $z$ is up.  A similar modification to the Riemann problem for this equation is made to incorporate fault slip.  In order to simplify the extension to three dimensions, only faults with strike angle of  $0$ degrees (top edge pointing north) and rake angle of $90$ degrees (slip is in dip direction) are considered.  Note these types of faults are the most common in subduction zones.  The mapped grid is an extension of the two-dimensional mapping near the fault: $\mathcal{M}(x,y,z) = \mathcal{M}^s(x,y,z)\phi(x,z)/z_c + \mathcal{M}^f (x,y,z)(1 - \phi(x,z)/z_c)$.  Note that this mapping is invariant to $y$, and thus the general Riemann solutions either involve the eigenvectors to $n_xA + n_zC$ or to $B$, because the normal to the cell-faces is either $n = [n_x,0,n_z]^T$ or $n=[0,1,0]^T$.
\par
If $n=[0,1,0]^T$, the standard Riemann solution for (\ref{eq:3delasticity}) is used because no slip is to be imposed.  Thus, assume $n = [n_x,0,n_z]^T$.  Denote the tangent vectors $\tau_1 = [n_z, 0, -n_x]^T$ and $\tau_2 = [0, 1, 0]^T$, noting that $\tau_1$ points in the rake direction and $\tau_2$ is orthogonal to $\tau_1$.  The eigenvectors of $n_xA + n_zC$ are
\begin{equation}\label{eq:waves3d}
	r^p_\pm = \bcm \lambda + 2\mu n_x^2\\ \lambda\\ \lambda + 2\mu n_z^2 \\ 0 \\ 0 \\ 2 \mu n_x n_z \\
					 \mp c_p n_x \\ 0  \\ \mp c_p n_z \ecm, \quad
	r^{s1}_\pm = \bcm -2\mu n_x n_z  \\ 0 \\  2\mu n_x n_z \\ 0 \\ 0 \\ \mu (n_z^2 - n_x^2)\\
					\pm c_s n_z \\ 0 \\ \mp c_s n_x \ecm, \quad \text{and }
	r^{s2}_\pm = \bcm 0 \\ 0 \\ 0 \\ \mu n_x \\ \mu n_z \\ 0\\
					0 \\ \mp c_s \\ 0 \ecm.
\end{equation}
The initial and resulting states are again related via these eigenvectors: $q_\pm = q^*_\pm \mp \alpha^p_\pm r^p_\pm \mp \alpha^{s1}_\pm r^{s1}_\pm \mp \alpha^{s2}_\pm r^{s2}_\pm$.  The conditions at the cell-face are now continuity of normal traction ($\sigma.n \cdot n$) and normal velocity ($u \cdot n$), an evenly imposed slip rate $s$ in the two $\tau_1$ velocities, and zero velocity in the $\tau_2$ direction.  As before, these conditions are summarized in matrix form:
\begin{equation*}
	\begin{split}
		&P_n q_- = P_n q_+, \quad P_{\tau_1} q_- = s/2, \quad P_{\tau_1} q_+ = -s/2, \quad \text{and } \ P_{\tau_2} q_\pm = 0,\\
		&\text{where } P_n q := \bcm n_x^2 & 0 & n_z^2 & 0 & 0 & 2n_x n_z & 0 & 0 & 0 \\ 0 & 0 & 0 & 0 & 0 & 0 & n_x & 0 & n_z\ecm q = \bcm \sigma _n \\ u_n \ecm, \\
		&\quad \quad P_{\tau_1} q :=  \bcm 0 & 0 & 0 & 0 & 0 & 0 & n_z & 0 & -n_y \ecm q = \bcm u_{\tau_1} \ecm, \\
		&\text{ and } P_{\tau_2} q := \bcm 0 & 0 & 0 & 0 & 0 & 0 & 0 & 1 & 0 \ecm q = \bcm u_{\tau_2} \ecm.
	\end{split}
\end{equation*}
The resulting states are again exchanged for the initial states and eigenvectors, and again it is useful to note that $P_n r^{s1}_\pm = P_n r^{s2}_\pm = 0$, $P_{\tau_1} r^p_\pm = P_{\tau_1} r^{s2}_\pm = 0$, and $P_{\tau_2} r^p_\pm = P_{\tau_2} r^{s1}_\pm = 0$.  The solution to the resulting system is
\begin{equation}\label{eq:rpsolution3d}
	\alpha^p_\pm = \frac{c_{p\mp} \Delta \sigma^*_n \mp B_\mp \Delta u^*_n}{c_{p+}B_- + c_{p-}B_+}, \quad
	\alpha^{s1}_\pm = \frac{u^*_{\tau_1} \pm s/2}{c_{s\pm}}, \quad \text{and }
	\alpha^{s2}_\pm = 0.
\end{equation}

\section{Comparison of Numerical Results with the Okada Solution}
\label{sec:comparison}

In Sect.~\ref{sec:2d}, fault slip is modeled in a plane-strain,  linearly elastic solid (\ref{eq:2delasticity}) by modifying the corresponding Riemann problems (\ref{eq:waves})-(\ref{eq:rpsolution}) in a novel way.  This approach is then extended to three dimensions (\ref{eq:3delasticity}) and the modified Riemann solutions (\ref{eq:waves3d})-(\ref{eq:rpsolution3d}) given.  To verify this approach, the Okada solution is used as implemented in the GeoClaw `dtopotools' Python module\footnote{See {\tt www.clawpack.org/okada.html}}.  Surface deformation is computed by numerically integrating the velocity at fixed gauge locations along the surface.  These values are then compared to those of the Okada solution at the same spatial locations.

\subsection{Numerical results in two dimensions}

\begin{figure}[t]
	\center
	\subfloat[$t=4$s]{
		\begin{minipage}{0.5\textwidth}
			\includegraphics[width=\textwidth]{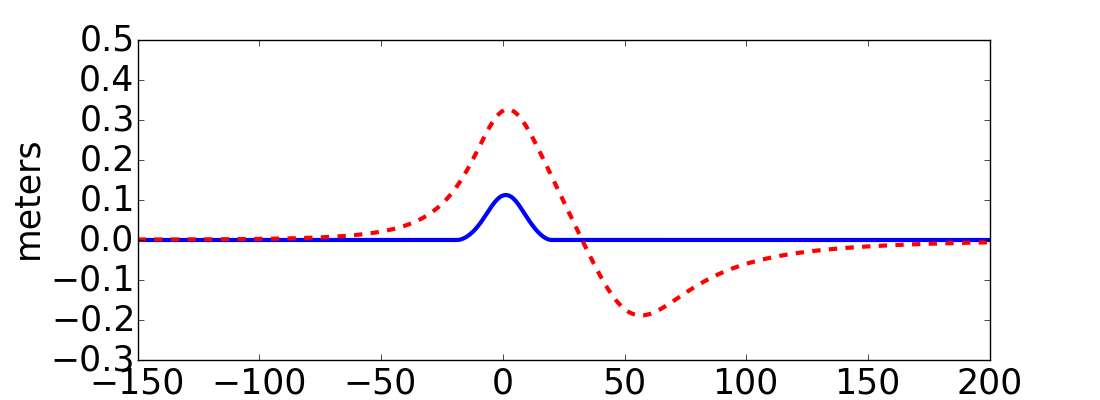} \\
    			\includegraphics[width=\textwidth]{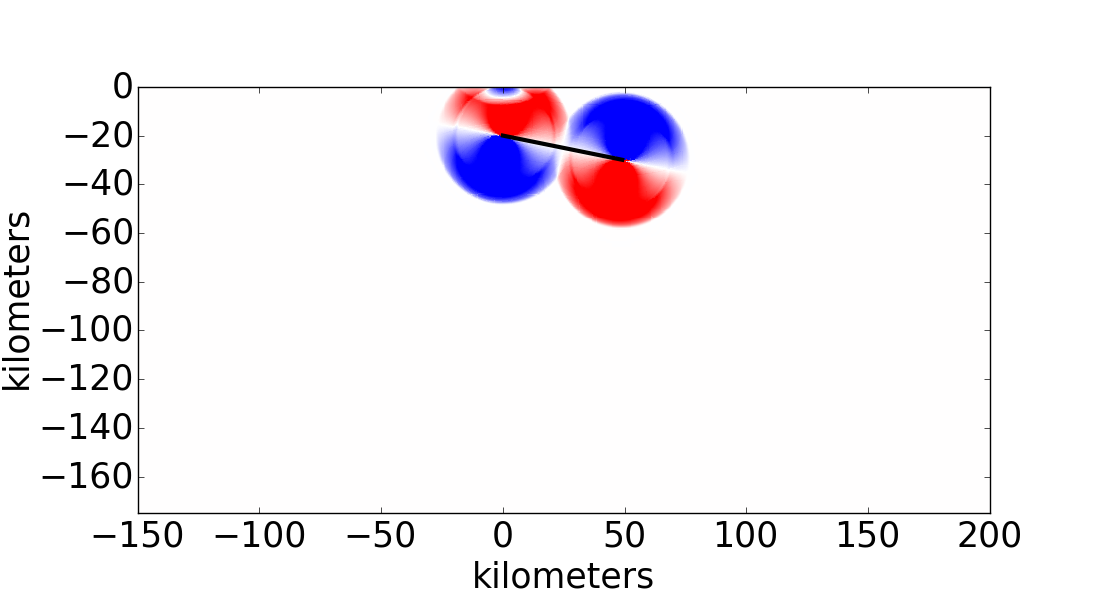}
    		\end{minipage}
    		}
	\subfloat[$t=10$s]{
		\begin{minipage}{0.5\textwidth}
			\includegraphics[width=\textwidth]{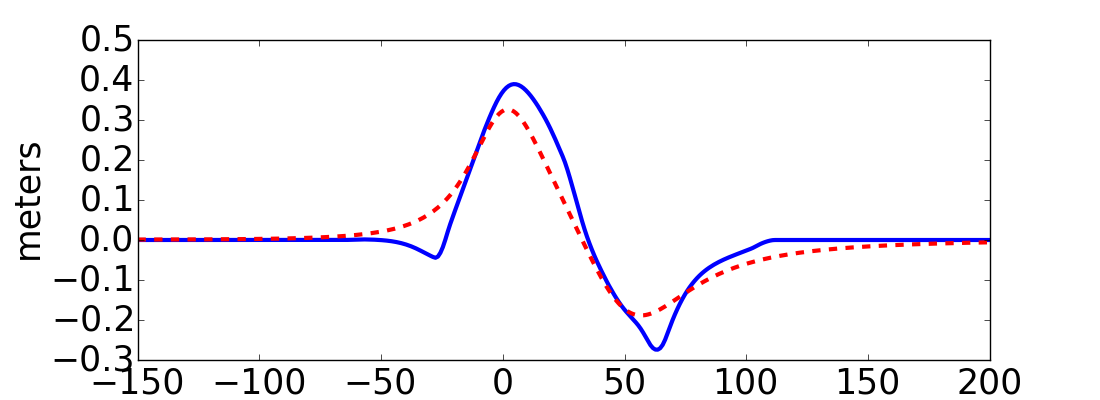} \\
    			\includegraphics[width=\textwidth]{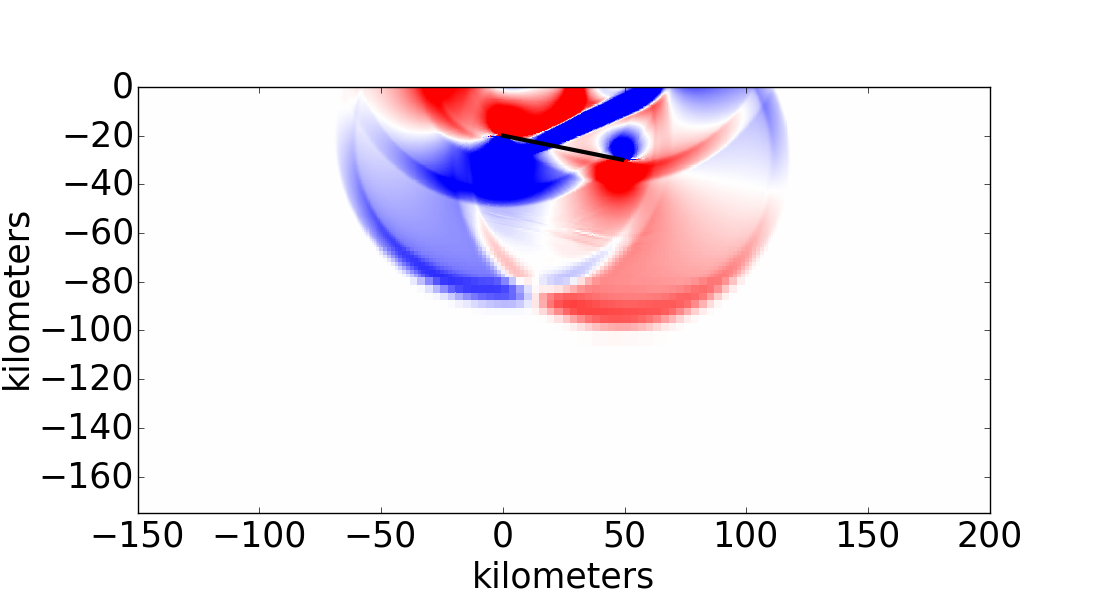}
    		\end{minipage}
    		}\\
	\subfloat[$t=20$s]{
		\begin{minipage}{0.5\textwidth}
			\includegraphics[width=\textwidth]{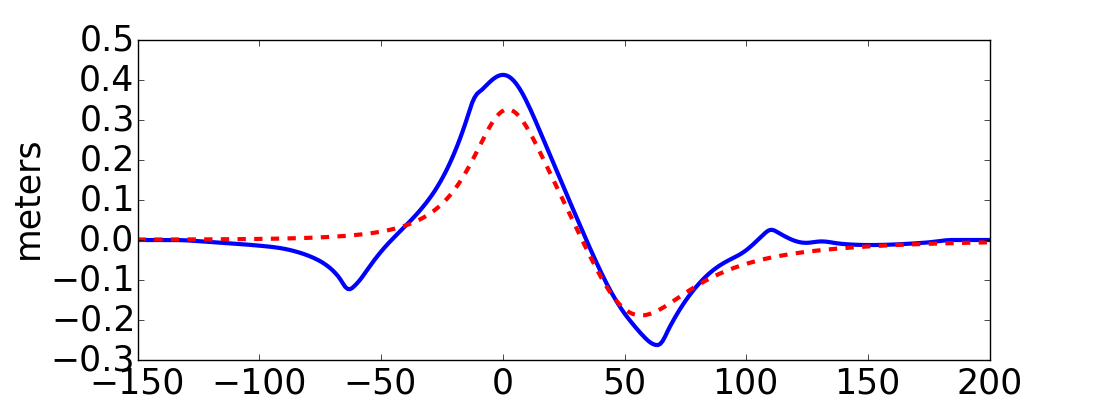} \\
    			\includegraphics[width=\textwidth]{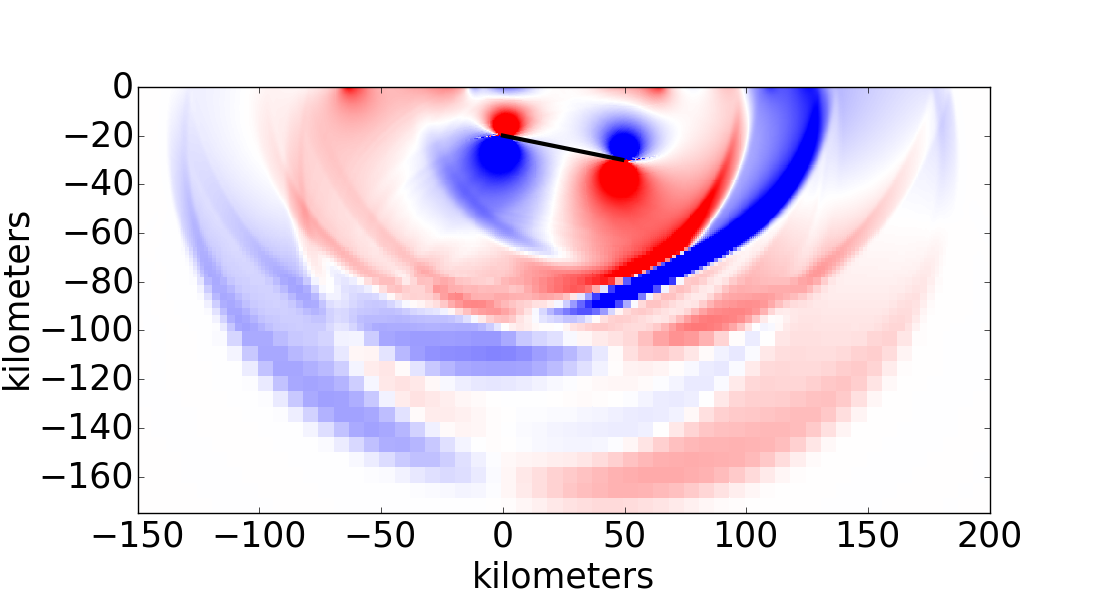}
    		\end{minipage}
    		}
	\subfloat[$t=30$s]{
		\begin{minipage}{0.5\textwidth}
			\includegraphics[width=\textwidth]{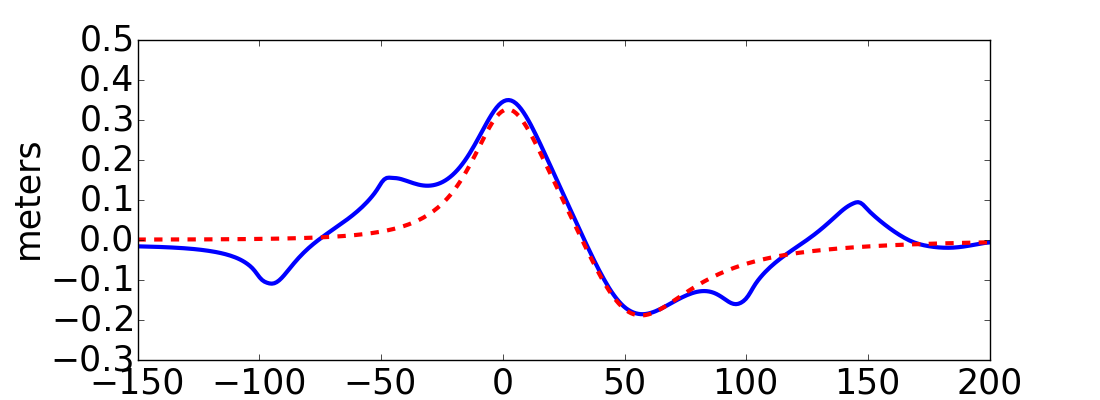} \\
    			\includegraphics[width=\textwidth]{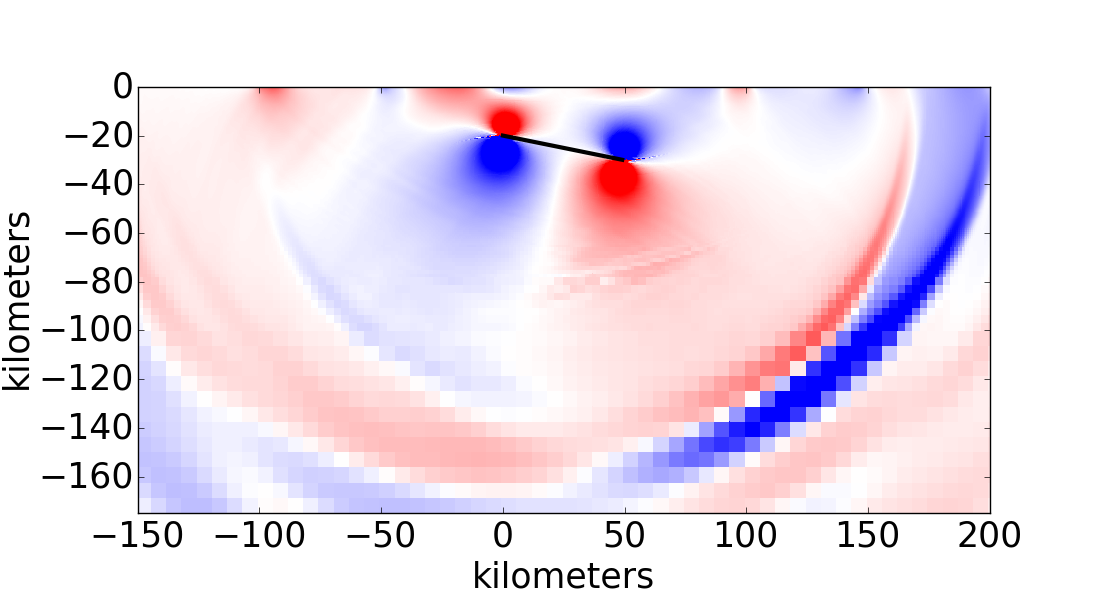}
    		\end{minipage}
    		}\\
	\subfloat[$t=40$s]{
		\begin{minipage}{0.5\textwidth}
			\includegraphics[width=\textwidth]{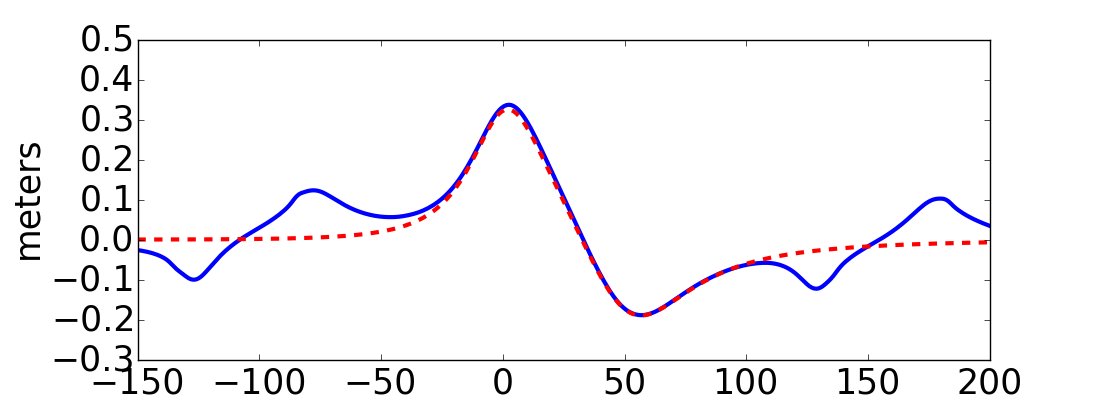} \\
    			\includegraphics[width=\textwidth]{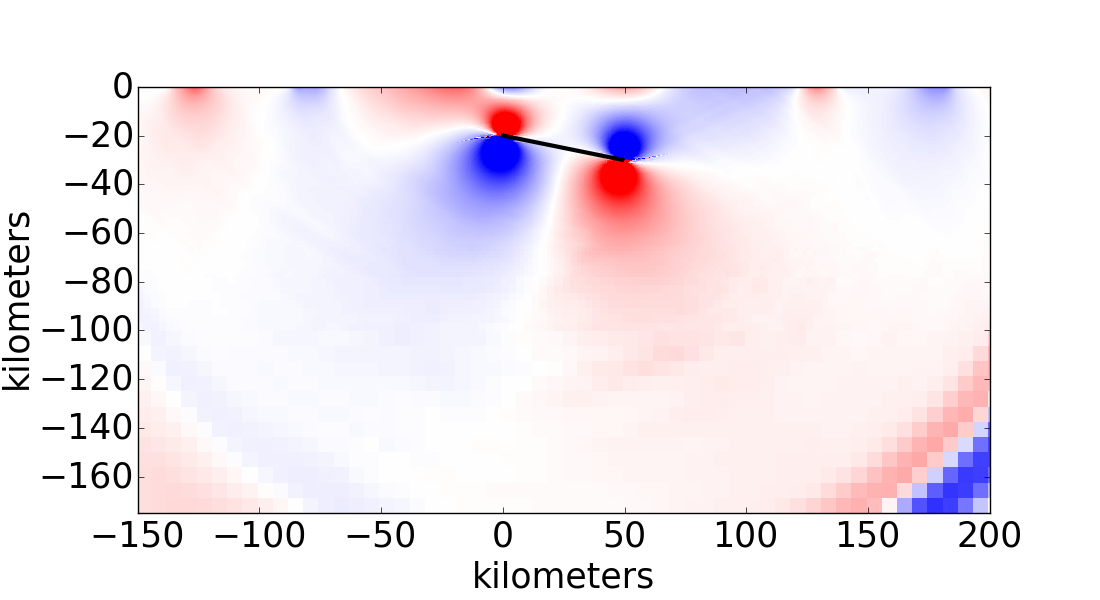}
    		\end{minipage}
    		}
	\subfloat[$t=70$s]{
		\begin{minipage}{0.5\textwidth}
			\includegraphics[width=\textwidth]{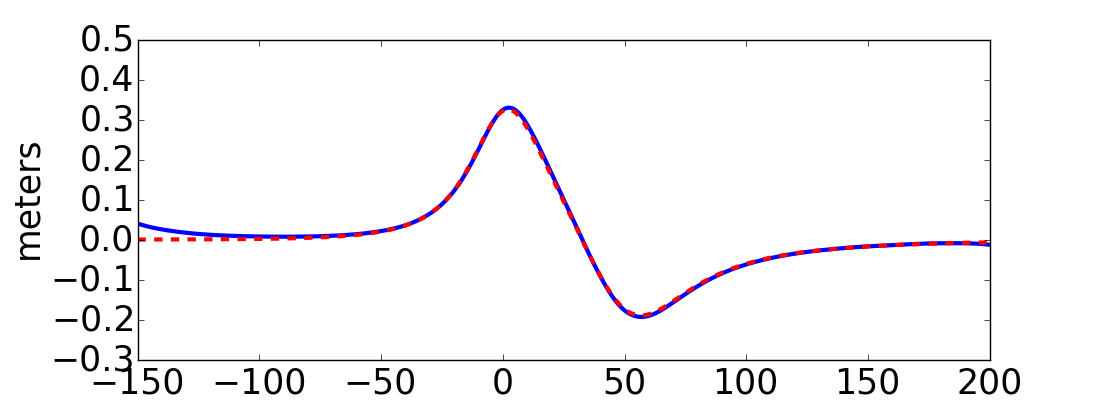} \\
    			\includegraphics[width=\textwidth]{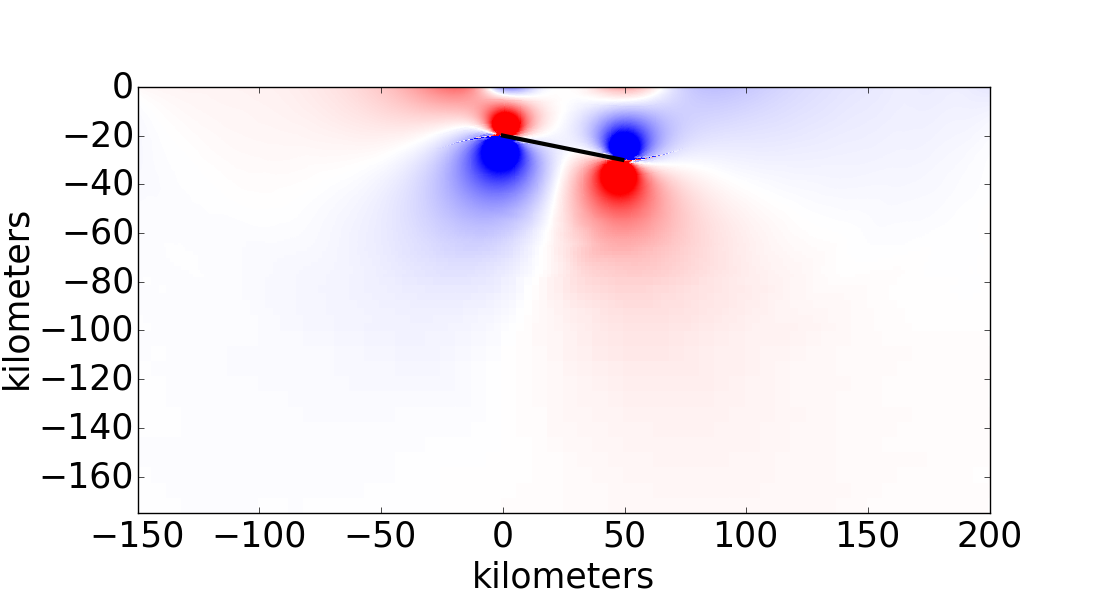}
    		\end{minipage}
    		}
	\caption[Compression/tension waves and resulting vertical surface displacement evolving to the Okada solution]{Results for a $1$km, $1$s uniform slip across a fault with dip $0.2$rad, top-edge depth $20$km, and width $50$km.  The upper plot at each time shows the surface deformation (\textcolor{blue}{blue solid line} - numerical surface displacement, \textcolor{red}{red dashed line} - Okada solution), and the lower plot shows the compression/tension waves (\textcolor{red}{red}: compression, \textcolor{blue}{blue}: tension).}
	\label{fig:faultandsurfacewaves}
\end{figure}

For the plane-strain case, a baseline fault is chosen with unit slip for $0 \le t \le 1$, dip $0.2$ radians ($\approx 11.5$ degrees), top-edge depth $20$km, and fault plane width $50$km.   Fig.~\ref{fig:faultandsurfacewaves} shows the resulting simulation at various times.
Note that the seismic waves interact with the free surface at the top of the domain, both reflecting off of and traveling along the surface, leaving behind a static deformation.  As the waves propagate away, the surface deformation approaches that of the Okada solution.  At $t=90$s, with the same AMR strategy used for the results in Fig.~\ref{fig:faultwaves}, the numerical solution was shown to converge to the Okada solution as the number of cells across the fault at the coarsest level is increased.  With $8$ cells, the difference between the numerical results at $t=90$s and the Okada solution is less than $2\%$ relative to the maximum Okada deformation.
\par
\begin{figure}[t]
	\center
	\subfloat[baseline parameters]{
		\begin{minipage}{0.32\textwidth}
			\includegraphics[width=\textwidth]{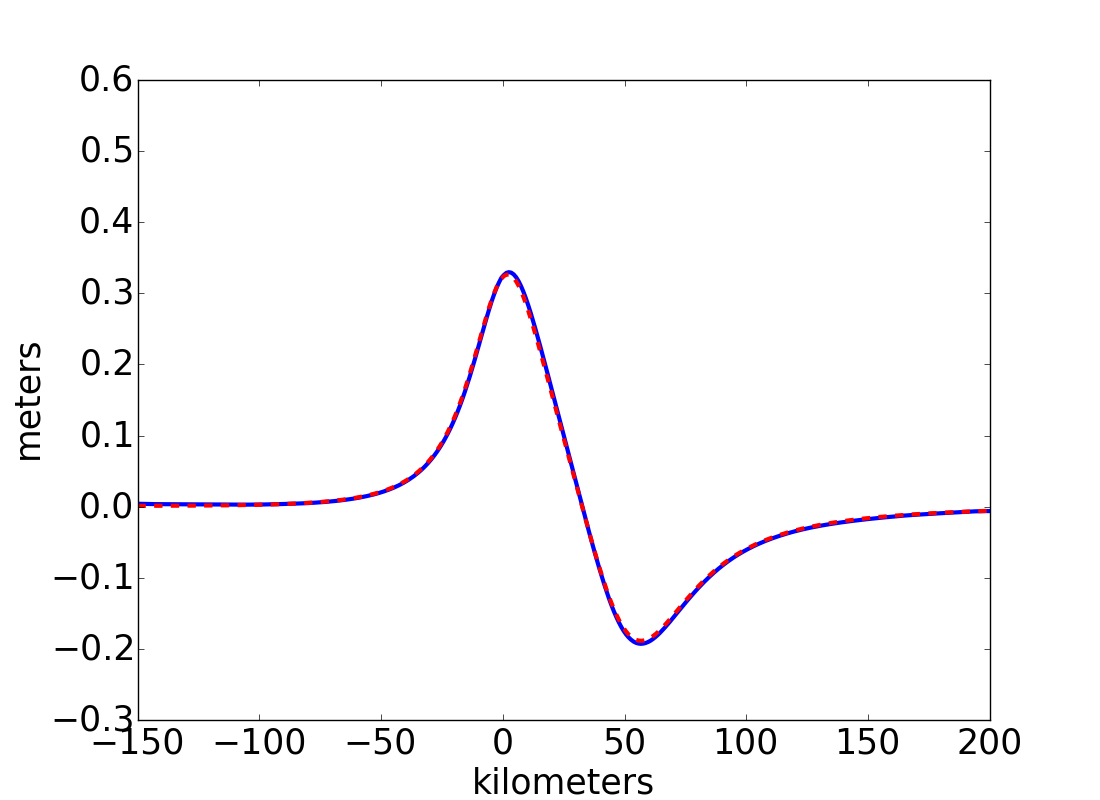}
			\includegraphics[width=\textwidth]{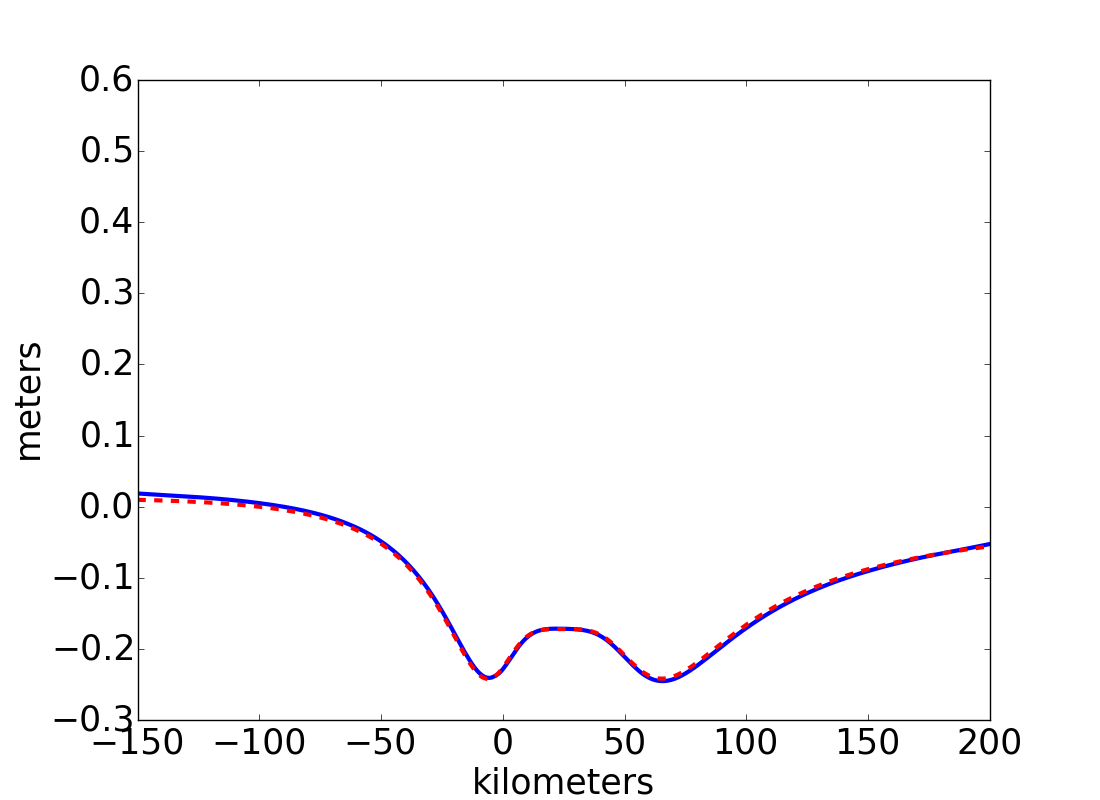}
		\end{minipage}
	}
	\subfloat[deeper top-edge $40$km]{
		\begin{minipage}{0.32\textwidth}
			\includegraphics[width=\textwidth]{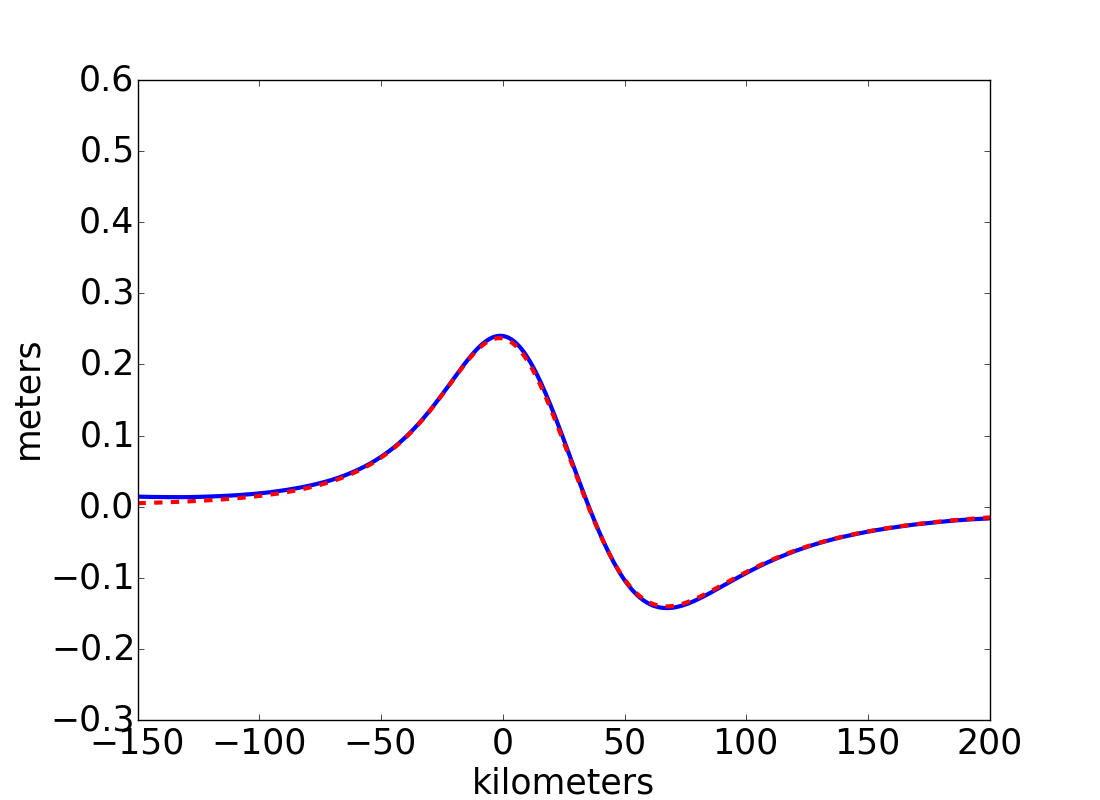}
			\includegraphics[width=\textwidth]{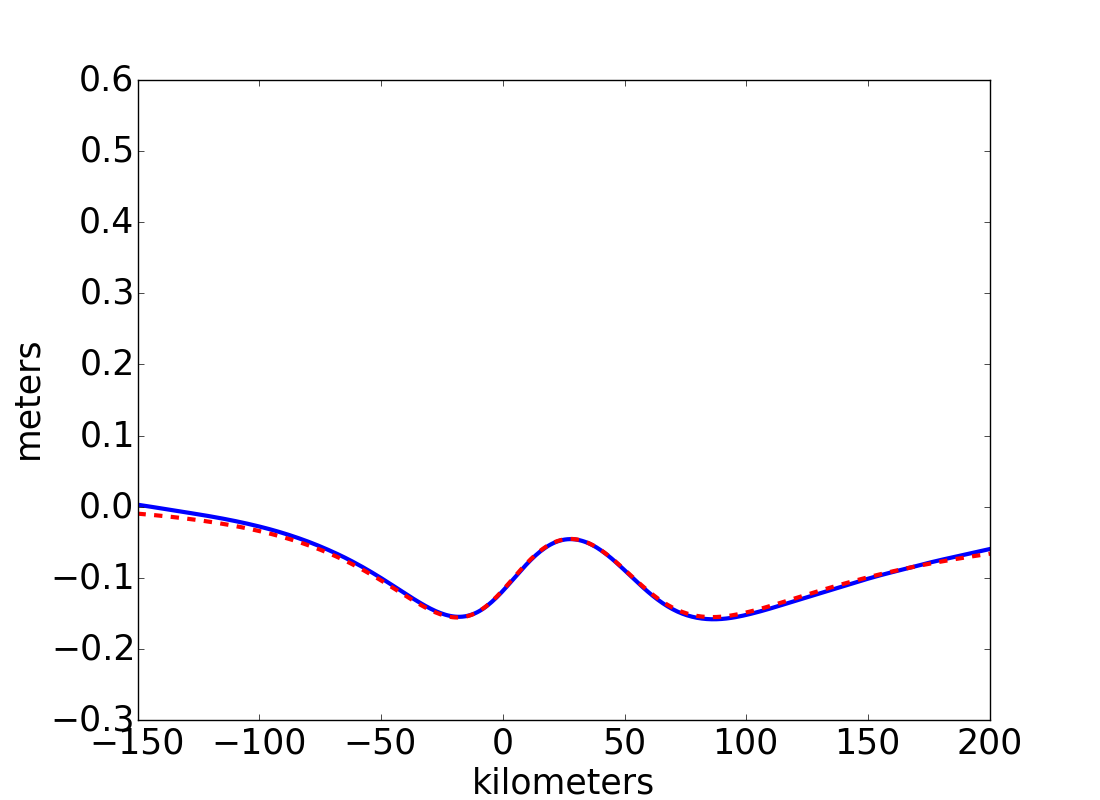}
		\end{minipage}
	}
	\subfloat[greater dip angle: $0.4$rad]{
		\begin{minipage}{0.32\textwidth}
			\includegraphics[width=\textwidth]{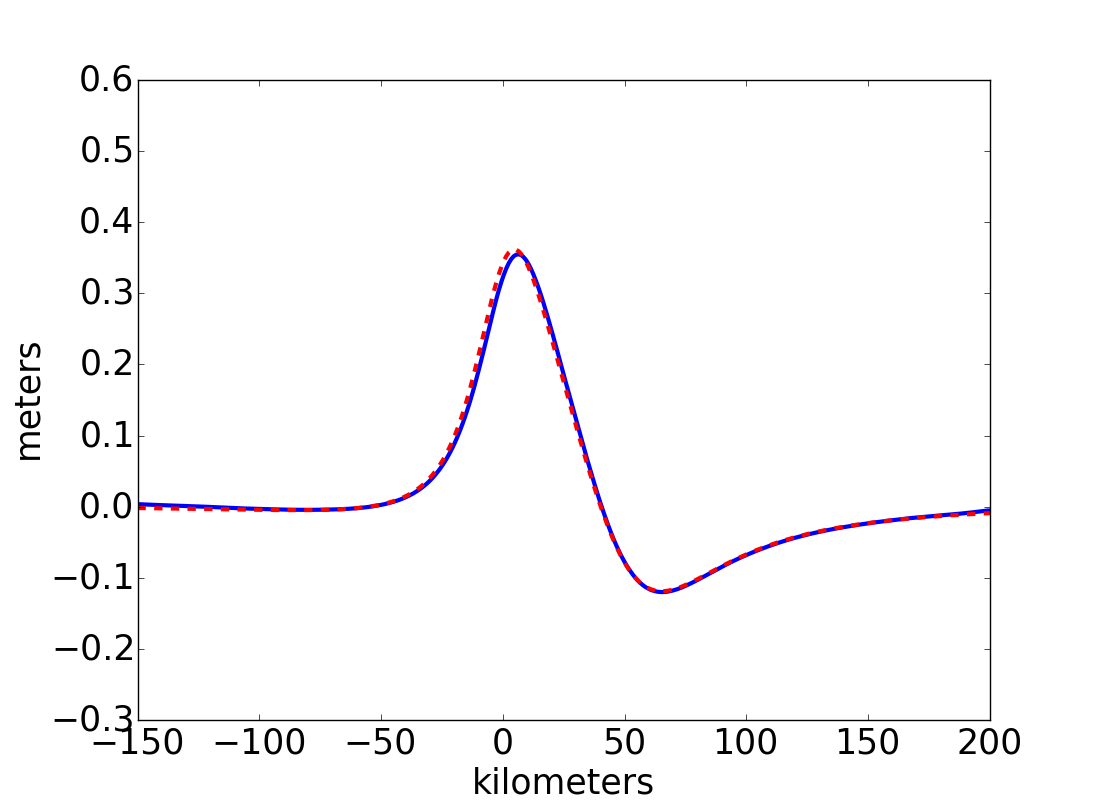}
			\includegraphics[width=\textwidth]{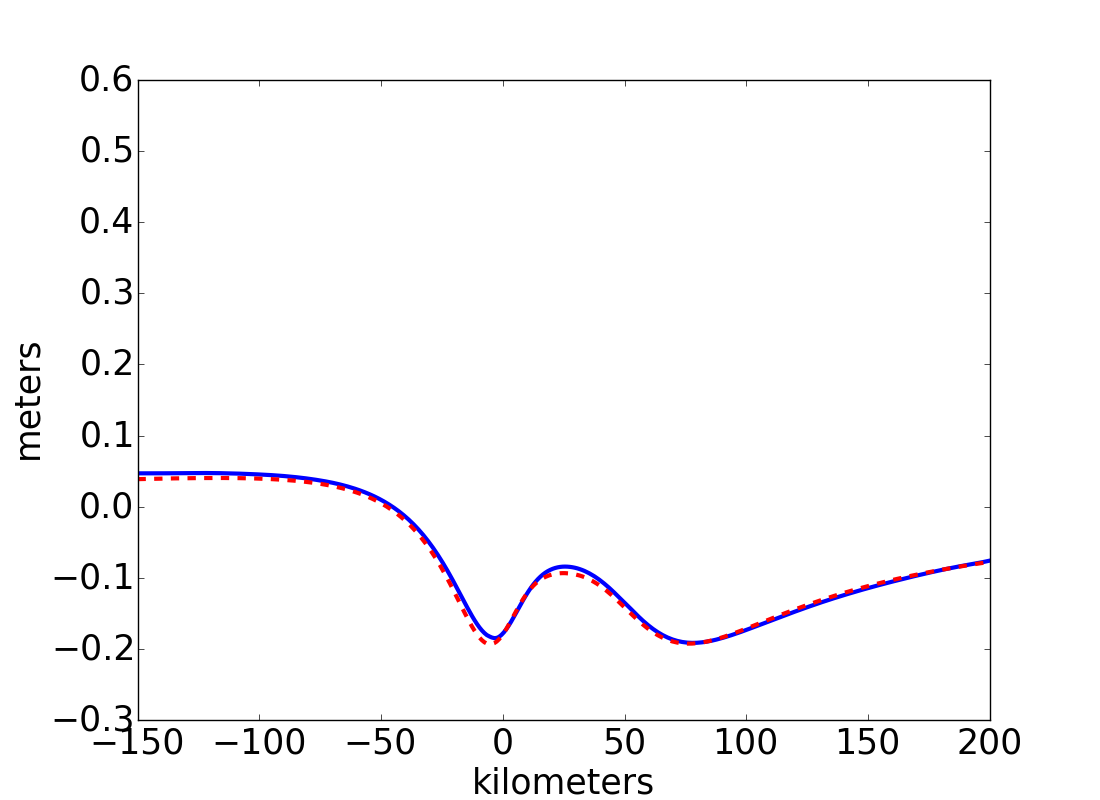}
		\end{minipage}
	}\\
	\subfloat[shorter fault width: $25$km]{
		\begin{minipage}{0.32\textwidth}
			\includegraphics[width=\textwidth]{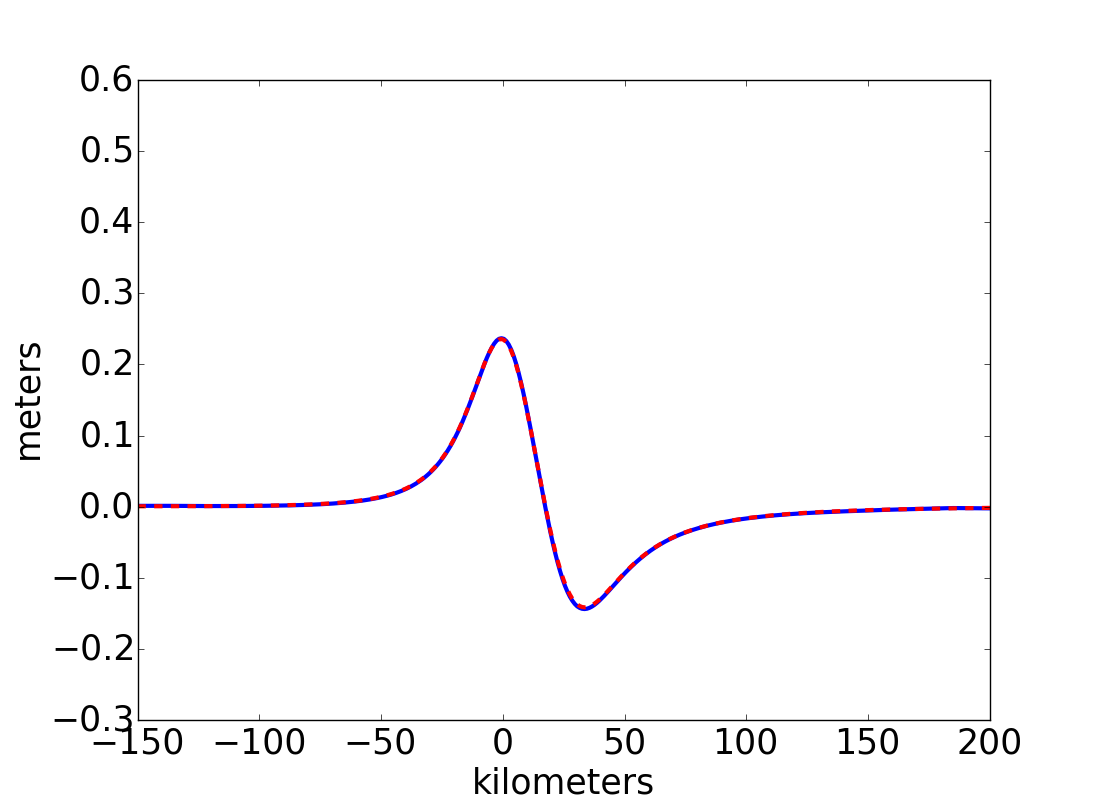}
			\includegraphics[width=\textwidth]{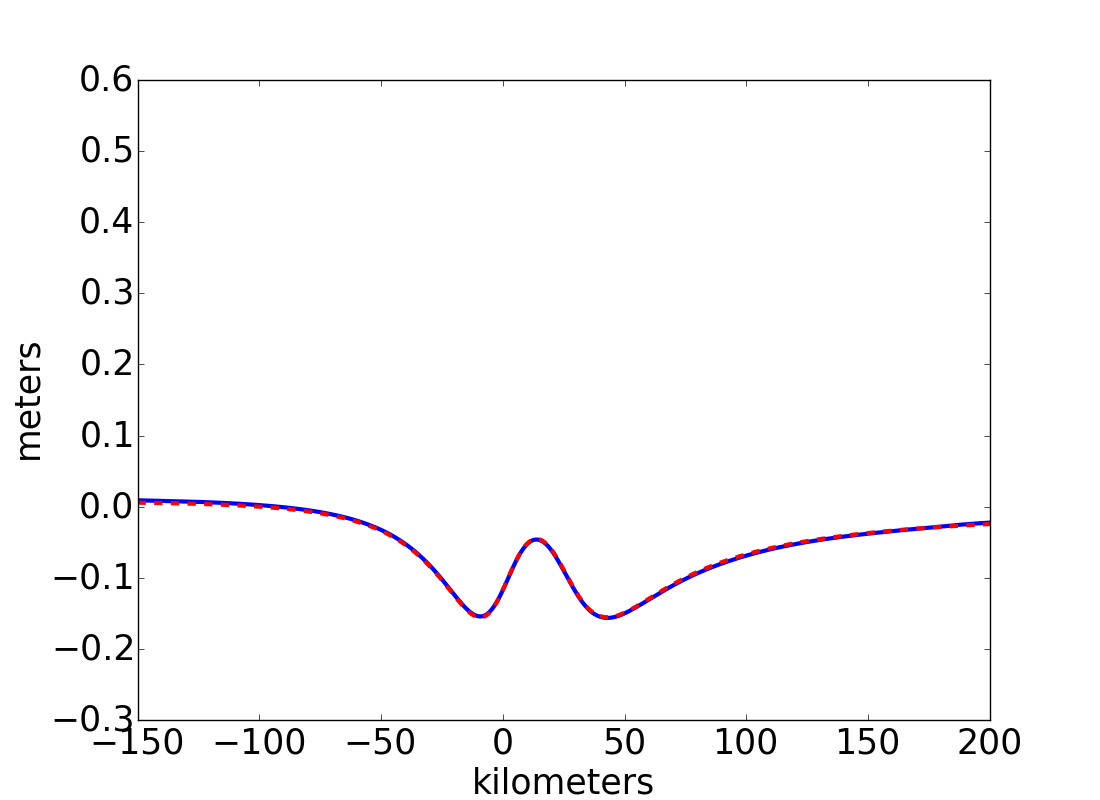}
		\end{minipage}
	}
	\subfloat[smooth slip profile]{
		\begin{minipage}{0.32\textwidth}
			\includegraphics[width=\textwidth]{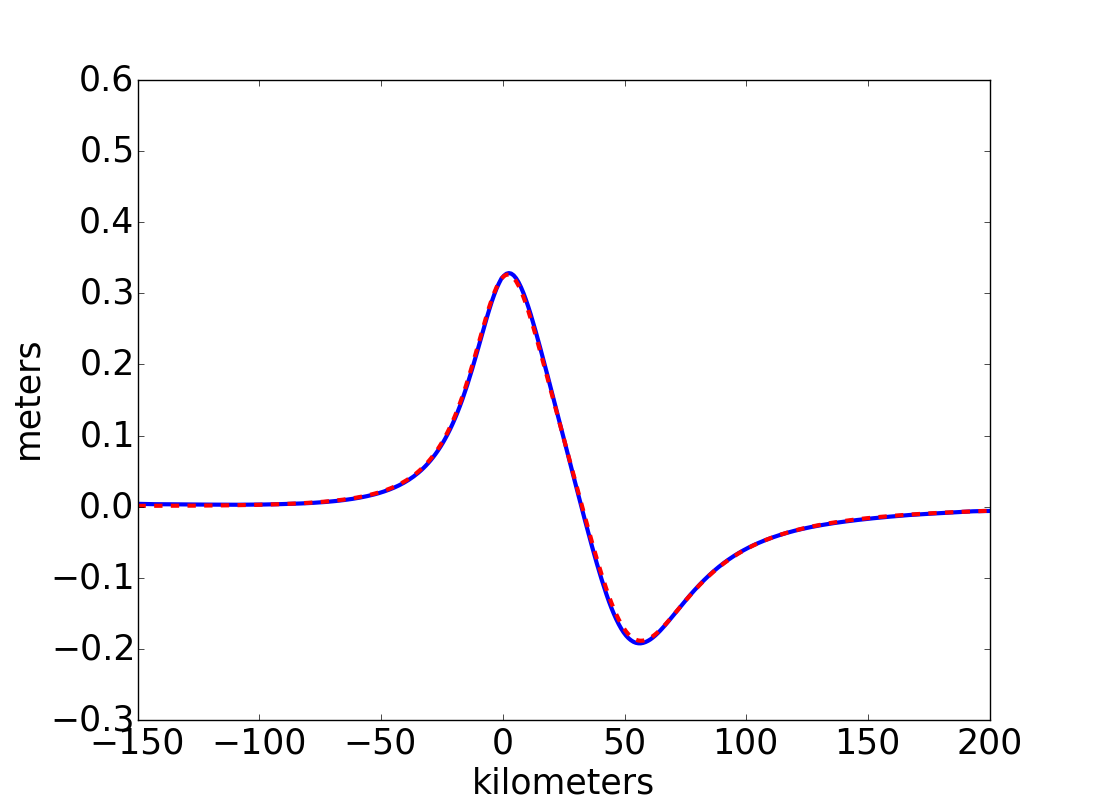}
			\includegraphics[width=\textwidth]{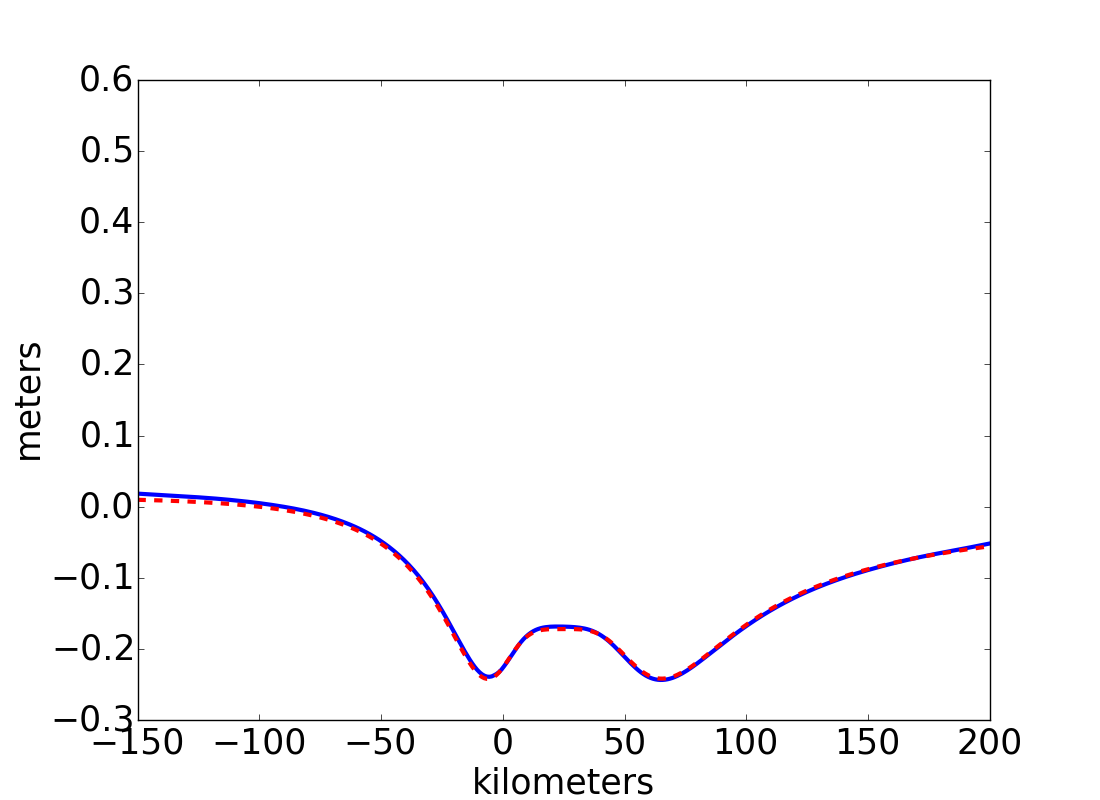}
		\end{minipage}
	}
	\subfloat[bimodal slip profile]{
		\begin{minipage}{0.32\textwidth}
			\includegraphics[width=\textwidth]{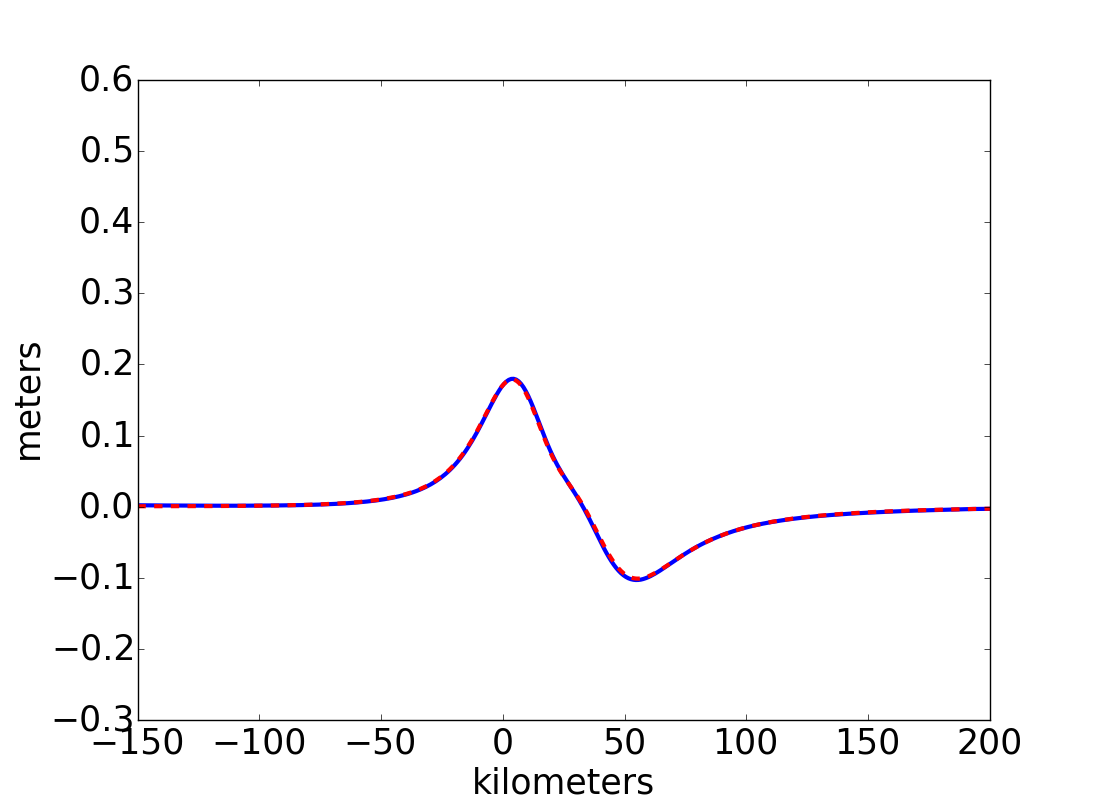}
			\includegraphics[width=\textwidth]{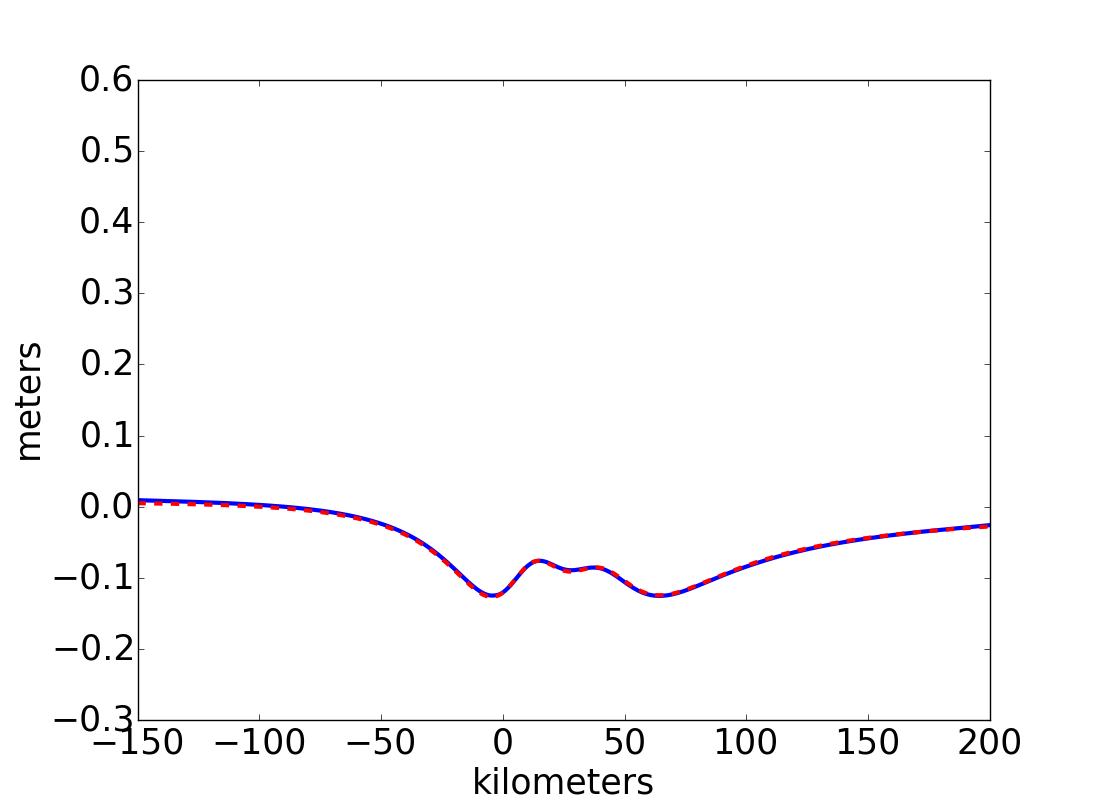}
		\end{minipage}
	}
	\caption[Vertical and horizontal surface displacement for various fault parameters matching that of Okada]{Vertical (top) and horizontal (bottom) surface displacement at $t=90$s matching that of Okada for fault parameters varying from the baseline of a $1$km, $1$s uniform slip across a fault with dip $0.2$rad, top-edge depth $100$km, and width $50$km \textcolor{blue}{Blue solid line}: numerical solution, \textcolor{red}{red dashed line}: Okada solution).}
	\label{fig:variousparameters}
\end{figure}

To ensure the robustness of this approach to physical fault parameters, the baseline fault is modified to obtain three additional faults.  Fig.~\ref{fig:variousparameters} shows the results for a fault with greater top-edge depth ($40$km vs $20$km), another with a greater dip angle ($\approx 22.9$ degrees vs $\approx 11.5$ degrees), and a third with shorter width ($25$km vs $50$km).
Note that all fault geometries studied show very close agreement, both in vertical and horizontal displacement, with the corresponding Okada solutions.  As an additional test, the fault slip profile is also varied.  Let $W$ denote the fault width, $p \in [0,W]$ denote the down-dip distance from the top of the fault, $p_0=W/2$, $p_1=W/4$, and $p_2=3W/4$.  A smooth fault profile of $S(p) = [1+\cos(\pi(p-p_0)/p_0)]/2$ is chosen, along with a bi-modal profile $S(p) = [1 + \cos(\pi(p-p_1)/p_1)]/2$ for $p \in [0,p_0]$ and $S(p) = [1 +
\cos(\pi(p-p_2)/p_1)]/2$ for $p \in (p_0,W]$.  The results, also in Fig.~\ref{fig:variousparameters}, show good agreement with the corresponding Okada solutions in all cases.
\par
\subsection{Numerical results in three dimensions}
Next, the extension to three space dimensions from Sect.~\ref{sec:3d} is verified.
Similar to what was done in the two-dimensional simulations, a three-dimensional fault is chosen with unit slip for $0 \le t \le 1$, strike $0$ degrees, rake $90$ degrees, dip $0.2$ radians ($\approx 11.5$), centroid depth $25$km, $50$km of width in the dip direction, and $25$km of length in the strike direction.  The AMR strategy uses $8$ cells across the dip direction and $4$ cells across the strike direction at the coarsest level, with 3 additional levels of refinement.  In two dimensions, the corresponding resolution gave a relative error of less than $4\%$.
\begin{figure}[t]
	\center
	\subfloat[$t=4$s]{
		\begin{minipage}{0.5\textwidth}
			\includegraphics[width=\textwidth]{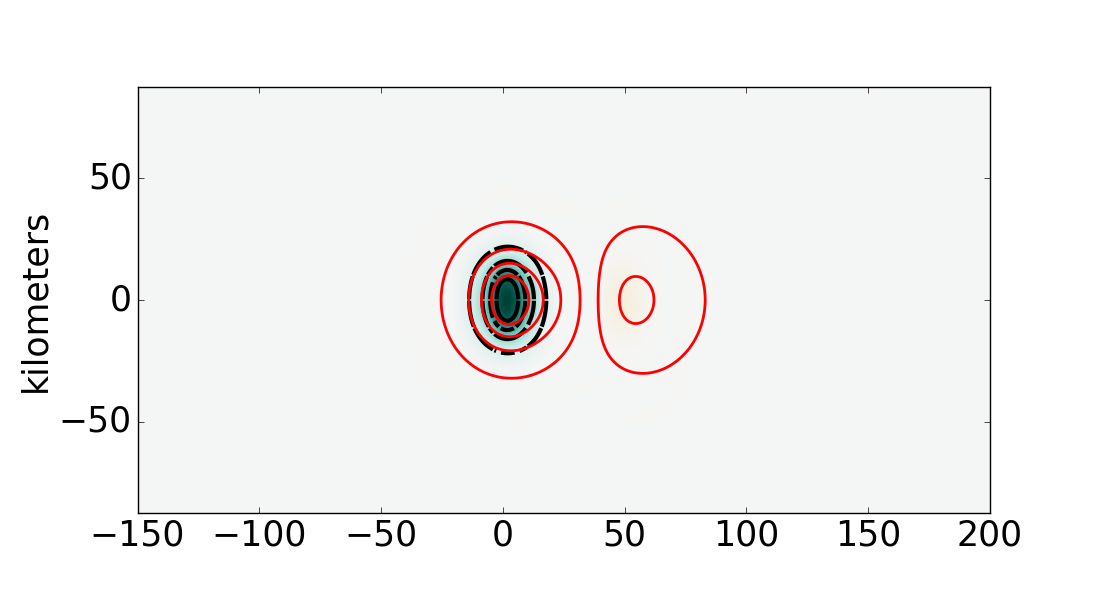} \\
    			\includegraphics[width=\textwidth]{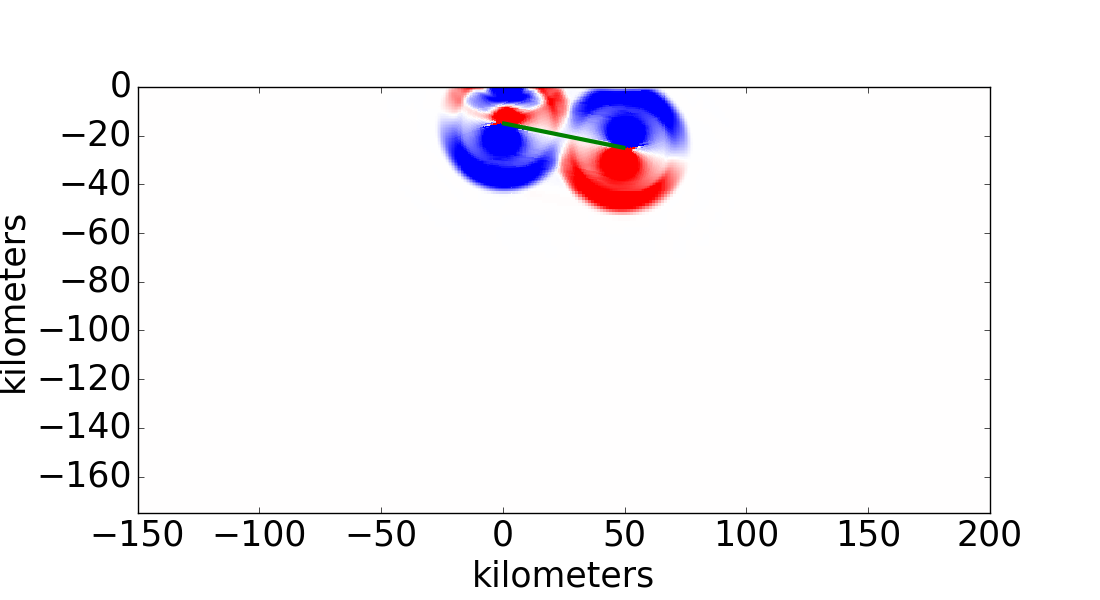}
    		\end{minipage}
    		}
	\subfloat[$t=10$s]{
		\begin{minipage}{0.5\textwidth}
			\includegraphics[width=\textwidth]{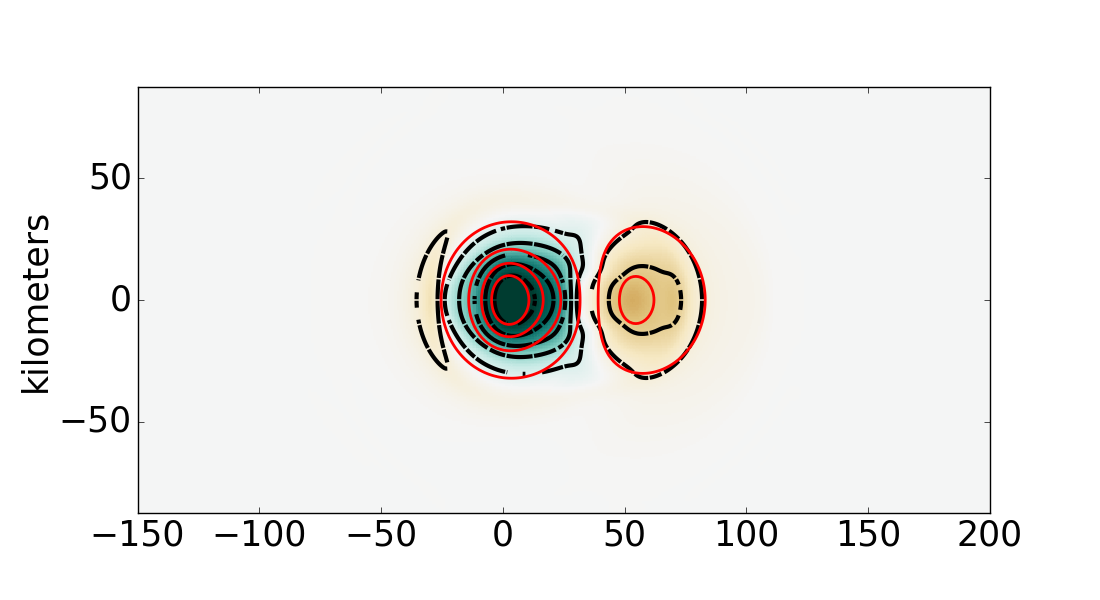} \\
    			\includegraphics[width=\textwidth]{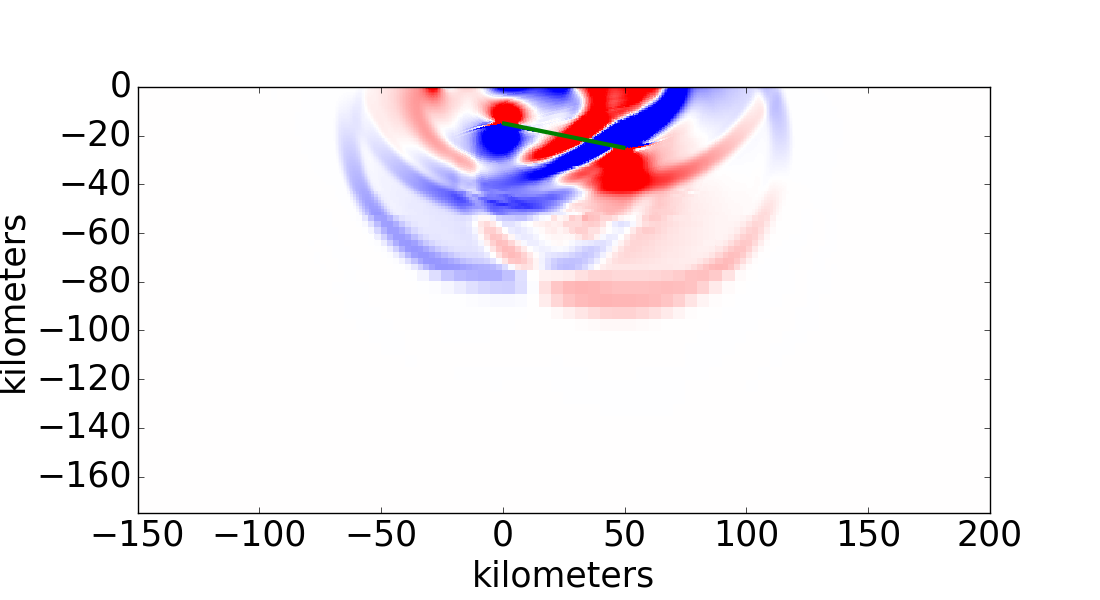}
    		\end{minipage}
    		}\\
	\subfloat[$t=20$s]{
		\begin{minipage}{0.5\textwidth}
			\includegraphics[width=\textwidth]{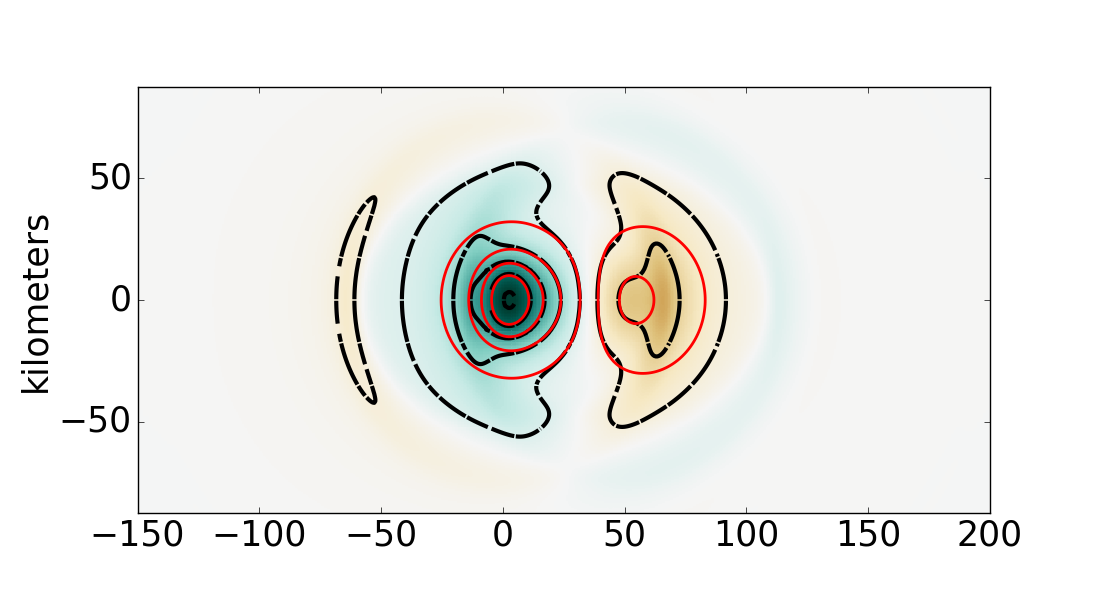} \\
    			\includegraphics[width=\textwidth]{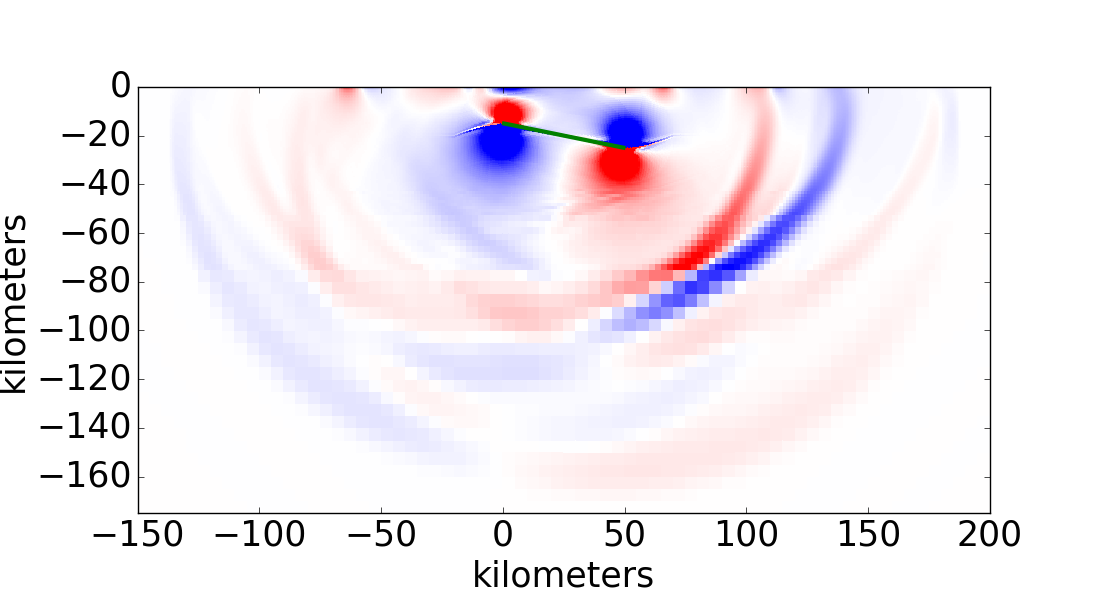}
    		\end{minipage}
    		}
	\subfloat[$t=30$s]{
		\begin{minipage}{0.5\textwidth}
			\includegraphics[width=\textwidth]{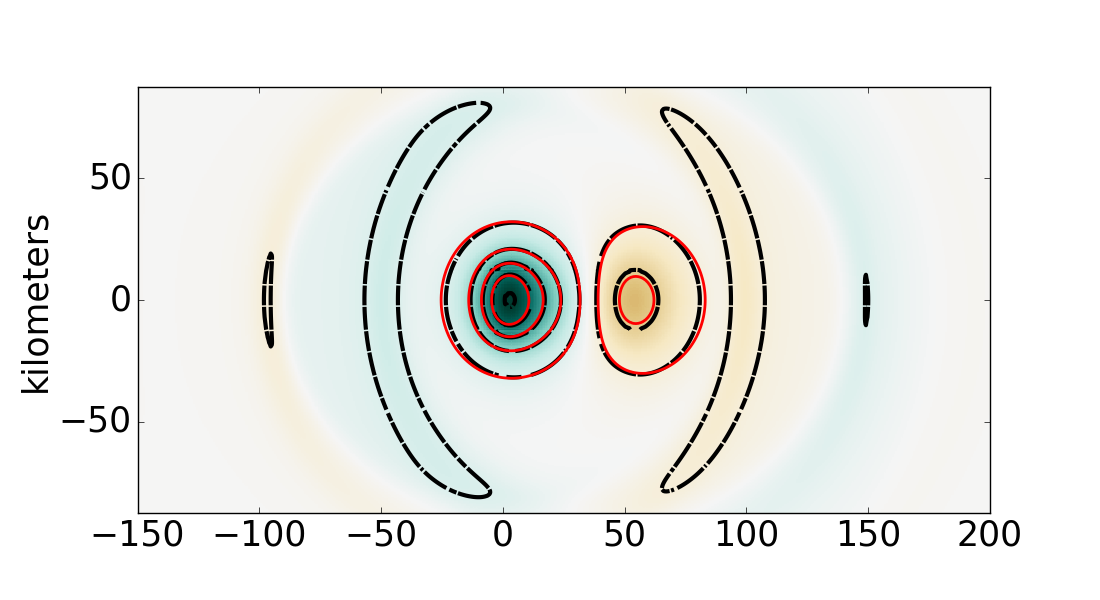} \\
    			\includegraphics[width=\textwidth]{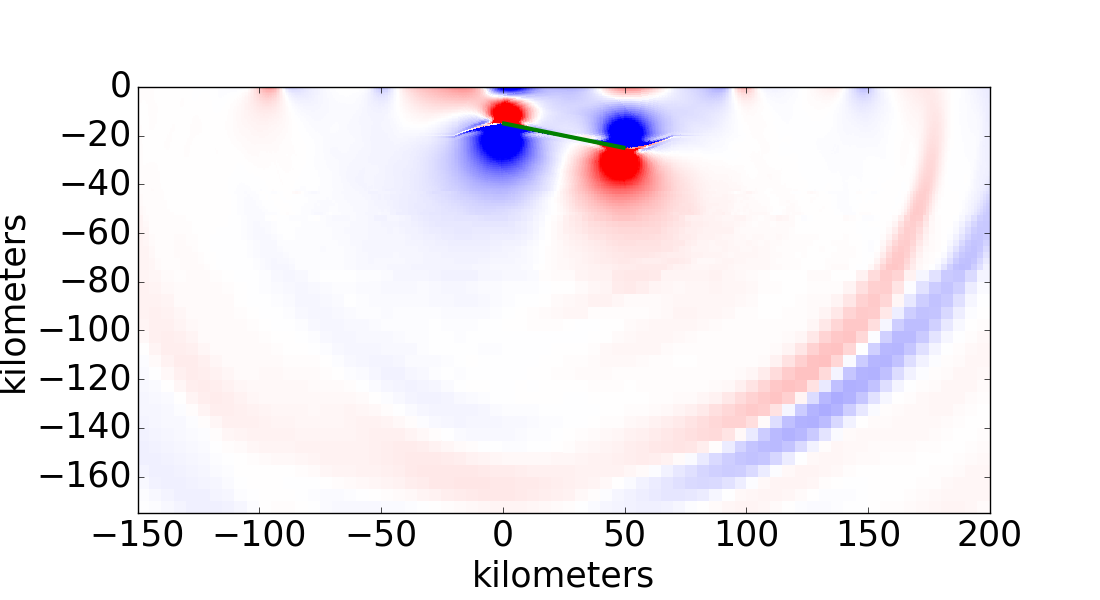}
    		\end{minipage}
    		}\\
	\subfloat[$t=40$s]{
		\begin{minipage}{0.5\textwidth}
			\includegraphics[width=\textwidth]{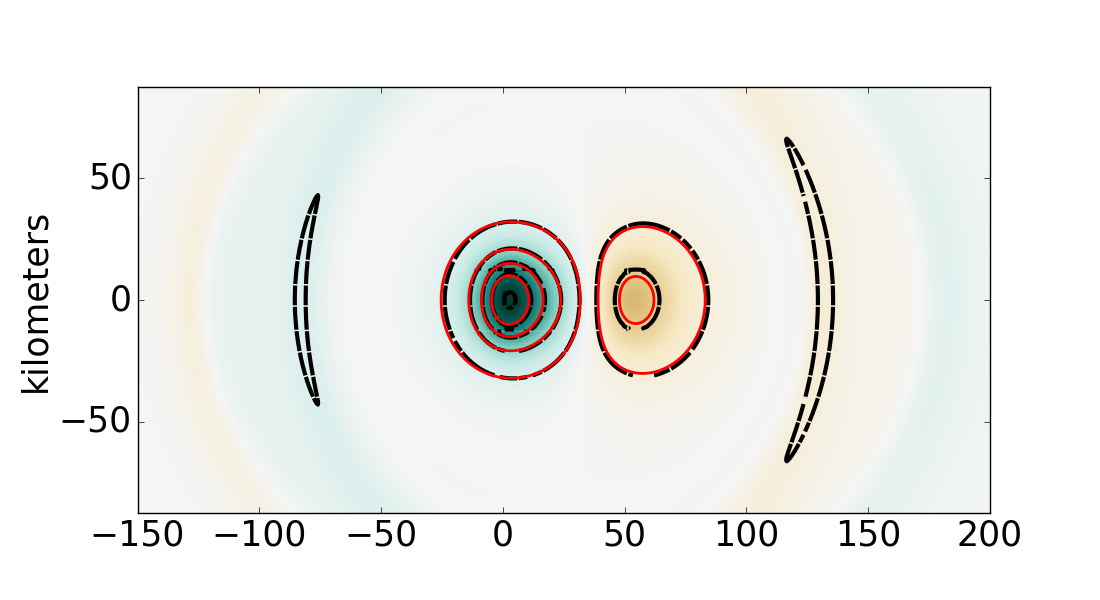} \\
    			\includegraphics[width=\textwidth]{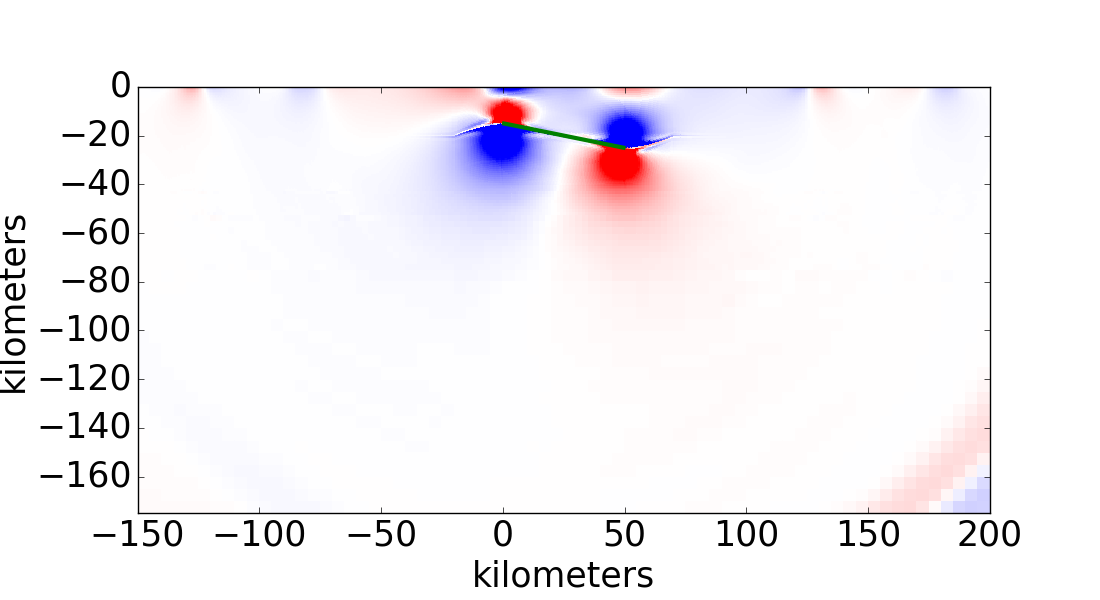}
    		\end{minipage}
    		}
	\subfloat[$t=70$s]{
		\begin{minipage}{0.5\textwidth}
			\includegraphics[width=\textwidth]{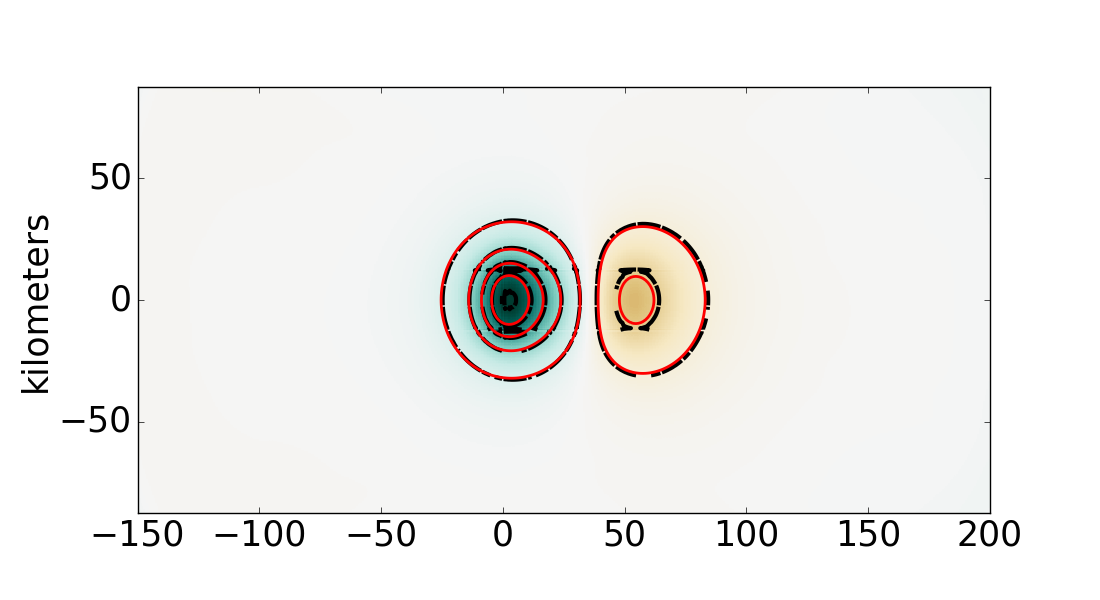} \\
    			\includegraphics[width=\textwidth]{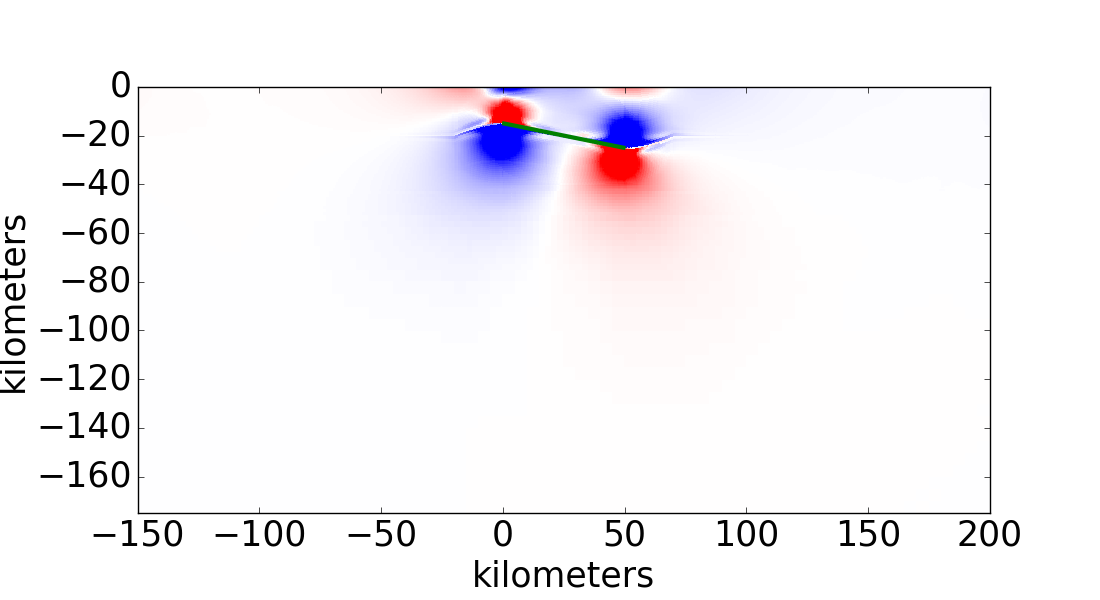}
    		\end{minipage}
    		}
	\caption[Vertical cross-section of compression/tension waves and plane view of surface deformation]{The upper plot at each time shows a plane view of the surface deformation (\textcolor{bluegreen}{blue-green}: uplift, \textcolor{brown}{brown}: subsidence, black lines: contours of numerical solution, \textcolor{red}{red lines}: contours of the Okada solution), and the lower plot shows a vertical cross-section of the compression/tension waves through the three-dimensional computation (\textcolor{red}{red}: compression, \textcolor{blue}{blue}: tension)
	}
	\label{fig:faultandsurfacewaves3d}
\end{figure}
Fig.~\ref{fig:faultandsurfacewaves3d} shows the seismic waves again interacting with the free surface at the top of the domain.  Again, this interaction causes deformation in the surface that, after the waves propagate out of the domain, approaches the Okada solution.  At $t=70$, the agreement between the numerical and Okada solution is very good.

\section{Conclusion and Discussion}
\label{sec:conclusion}

While the Okada solution is often adequate for tsunami simulation, the Okada assumptions of instantaneous displacement in a homogeneous half-space may be inadequate for some applications.  This work presents a novel way to introduce fault slip via the Riemann problems in the Clawpack simulation software to fully model the seismic waves and transient motion of the seafloor.  When solving the linear elasticity hyperbolic system, both in three dimensions  (\ref{eq:3delasticity}) and in the case of plane-strain (\ref{eq:2delasticity}), the Riemann solution at a cell interface imposes the slip velocity in the dip direction instead of imposing continuity of tangential traction and velocity.  The corresponding eigenvectors (\ref{eq:waves}),(\ref{eq:waves3d}) and coefficients (\ref{eq:rpsolution}),(\ref{eq:rpsolution3d}) are then used by the high-resolution wave-propagation algorithm implemented in Clawpack to solve for surface deformation.
\par
Two-dimensional results from a baseline fault initially differ from that of Okada for small $t$ but then evolve towards the steady-state solution for larger $t$, as is expected.  Physical parameters are varied to observe results for deeper faults, steeper faults, faults of various width, and faults with spatially varying slip profiles.  All the dynamic simulations show agreement with the corresponding Okada solution once the seismic waves propagate sufficiently away from the region of interest.  The approach is extended to three dimensions and applied to a dipping fault.  Again, the dynamic solution evolves towards the corresponding Okada solution.
\par
The Riemann solution presented here does not assume the elastic medium is homogeneous.  Thus, current work involves including an ocean layer as a second elastic medium with a zero shear modulus.  This permits direct observation of the sea surface deformation instead of relying on the assumption that the sea surface moves instantaneously with the seafloor.  Other current work involves incorporating varying topography, which can also be implemented via the mapped grid approach presented here.  Potential future work may include the varying densities of the subduction zone plates and a gravitational force.  Finally, it should be noted that this approach to modifying the Riemann problems can be generalized to include a variety of conditions on an interface embedded in an elastic solid, including jumps in tangential or normal traction.


\section*{acknowledgements}
The authors are grateful to Grady Lemoine for discussions of this work, in  particular those involving the modification of the Riemann problems to incorporate fault slip.

\bibliographystyle{siamplain}
\bibliography{references}

\end{document}